\documentclass[%
 reprint,
superscriptaddress,
 amsmath,amssymb,
 aps,
prc
]{revtex4-1}

\usepackage{graphicx}
\usepackage{dcolumn}
\usepackage{bm}
\allowdisplaybreaks

\begin{document}

\preprint{APS/123-QED}

\title{Optogenetic control of intracellular flows and cell migration:\\
a comprehensive mathematical analysis with a minimal active gel model}

\author{Oliver~M.~Drozdowski}
\author{Falko~Ziebert}%
\author{Ulrich~S.~Schwarz}%
 \email{Corresponding author: schwarz@thphys.uni-heidelberg.de}
\affiliation{Institute for Theoretical Physics, Heidelberg University, Philosophenweg 19, 69120 Heidelberg, Germany}
\affiliation{BioQuant, Heidelberg University, Im Neuenheimer Feld 267, 69120 Heidelberg, Germany}%
\date{\today}

\begin{abstract}
The actin cytoskeleton of cells is in continuous motion due to both polymerization of new filaments and their contraction by myosin II molecular motors. 
Through adhesion to the substrate, such intracellular flow can be converted into cell migration. 
Recently, optogenetics has emerged as a new powerful experimental method to control both actin polymerization and myosin II contraction. 
While optogenetic control of polymerization can initiate cell migration by generating protrusion, it is less clear if and how optogenetic control of contraction can also affect cell migration. 
Here we analyze the latter situation using a minimal variant of active gel theory into which we include optogenetic activation as a spatiotemporally constrained perturbation. 
The model can describe the symmetrical flow of the actomyosin system observed in optogenetic experiments, but not the long-lasting polarization required for cell migration. 
Motile solutions become possible if cytoskeletal polymerization is included through the boundary conditions. 
Optogenetic activation of contraction can then initiate locomotion in a symmetrically spreading cell and strengthen motility in an asymmetrically polymerizing one. 
If designed appropriately, it can also arrest motility even for protrusive boundaries.
\end{abstract}

\pacs{Valid PACS appear here}
\maketitle

\section{\label{sec:Introduction}Introduction}

Despite the large variety of different cell types in our body, all of them have
the ability to migrate. Migration is
essential for all cells in the developing embryo, and later is used by specific 
cell types for certain functions, such as white blood cells chasing intruders or
epithelial cells closing wounds \cite{ridley_cell_2003}. Very importantly, virtually all body cells can
revert back to the migratory mode, which is especially dangerous in the context
of cancer metastasis \cite{Friedl_2003_Nature_review_tumour_cell_invasion_migration,shatkin_computational_2020}. 
Moreover, there is a growing interest in the bottom-up construction of synthetic cells \cite{Goepfrich_2018_TendsBioTech_Bottom_up_construction_synth_cells},
but an understanding of the minimal ingredients for cell migration is still missing.

Although animal cells are complex systems and can use very different migration strategies,
the main physical basis of their migration capacity has been identified to be 
flow in the actomyosin cytoskeleton \cite{ridley_life_2011,blanchoin_actin_2014,bodor_cell_2020}.
Due to controlled assembly and disassembly of actin filaments, all major actin architectures in the cell
(lamellipodia, filopodia, lamellae, cortex and stress fibers) are continuously flowing.
When combined with adhesion to the substrate, this intracellular flow can be converted 
into productive cell migration, similar to the function of a automotive clutch \cite{elosegui-artola_control_2018}. 
For example, recently it has been shown through cell migration experiments on one-dimensional lanes and
theoretical modeling that increased flow leads to faster and more persistent cell migration \cite{maiuri_actin_2015}.
However, before cell migration is established as a steady state of the system, the cell first
has to polarize, either spontaneously or guided by some external cues.
A well established model system for this essential process is the keratocyte, a cell type
which lives on two-dimensional surfaces like the cornea of eyes and migrates with
a very steady shape. For these cells, it has been shown that they (and also their fragments)
can transition from a non-polarized stationary into a polarized motile state by application
of a simple mechanical perturbation \cite{Verkhovsky_1999_CurrBiol_Selfpolarization_directional_motility, Mogilner_2020_SemCellDevBio_Review_Keratocytes_Cell_Motility}. 

The spontaneous symmetry break in polarization underlying cell migration has attracted
considerable interest from theory, also because it resonates with symmetry breaking
transitions in other parts of physics, e.g. in spin systems or particle physics. 
The symmetry break in the actomyosin system underlying cell migration has been studied theoretically from 
different starting points, including actin polymerization \cite{Mogilner_2002_BioPhysJ_Actin_dynamics_rapidly_moving_cells, 
Kozlov_2007_BioPhysJ_polarization_bistability_cell_fragments, Fuhrmann_2007_JTheoBio_Initiation_cytoskeletal_asymmetry,Ziebert_2012_JRSInterface_model_self_polarization_phase_field}, myosin motor protein contraction \cite{Hawkins_2011_BioPhysJ_Spontaneous_contractility_mediated_flow, Rubinstein_2009_BioPhysJ_Viscpelastic_flow_keratocyte_lemellipod, Recho_2015_JMPS_Mechanics_motility_initiation_arrest} 
and cellular adhesion to substrates \cite{Banerjee_2011_EPL_Substrate_polarizes_active_gels,  Ziebert_2013_PLOS1_Adhesion_substrate_shape_motility_crawling}. 
These physics-based analyses have been complemented by mathematical
analysis of the reaction-diffusion equations that describe the signaling
networks that control actomyosin flow inside cells \cite{Mori_2008_BioPhysJ_Wave_pinning_cell_polarity_reaction_diffusion, Mori_2011_JApllMath_Asymptotic_bifurcation_analysis_wave_pinning, Jilkine_2011_PLOS_Comparison_models_cell_polarization_cues}.

On the experimental side, 
gaining a better understanding of the mechanisms underlying cell migration traditionally had to rely
on controling cell behaviour through genetic or biochemical means. Recently, however, 
optogenetics for the cytoskeleton has been introduced as a new and powerful tool to experimentally
control the protrusive and contractile activity of cells in a temporally and spatially controlled manner 
\cite{weitzman_optogenetic_2014,guglielmi_optogenetic_2016,Wittman_2020_CurrOpCelBio_Review_Optogenetic_Control_cell_dyn}. 
In this context, one usually engineers a light-sensitive construct into the cell that activates a central
regulator for the process of interest, e.g. the G-proteins Rac1 for actin polymerization and
RhoA for myosin II contractility, respectively. Is is easy to understand that local control of 
actin polymerization leads to directed protrusions and therefore cell migration, as demonstrated experimentally
\cite{wu_genetically_2009,wang_light-mediated_2010,valon_predictive_2015,Kato_2015_PLOS1_Lamellipodial_extension_optogenetics}.
In addition this strategy has already been applied to break the symmetry of actin-containing synthetic cells \cite{Jahnke_Spatz_2020_AdvBioSys_Engineer_light_responsive_actomyosin}.
However, it is less clear how optogenetic control of contractility could lead to cell migration.
Until now, optogenetic activation of myosin II contractility in single cells has been shown to lead to increased traction forces and
intracellular flow \cite{Oakes_2017_NatComm_Optogenetic_RhoA, Valon_2017_Nature_Optogenetic_control_cellular_forces},  but it has not been used yet to control cell migration,
although the level of myosin II activity is in fact known to influence the velocity of motile cells \cite{Even-Ram_2007_NatCellBiol_myosin_regulates_motility, Wilson_2010_Nat_myosin_activity_treadmilling, Doyle_2012_JCellSci_micro_evn_control_cell_migration, Barbier_2019_FrontImmun_myosin_migration_microenv}.

Here we address the question if and how optogenetics can be used to 
control cell migration from a theoretical point of view. A natural framework
to mathematically analyze this situation is the so-called active gel theory \cite{Kruse_2005_EPJE_Theory_Active_Polar_Gels_Paradigm_Cytoskeleton,Juelicher_2007_PhyRep_Review_Active_Behavior_Cytoskeleton,Prost_2015_Nature_Review_active_gel_physics}.
Here we use its simplest variant, which does not consider local polarization of the cytoskeleton,
but only its velocity field driven by local active stresses. Such a hydrodynamic system is strongly determined by
its boundary conditions and in general one can distinguish between two approaches when modeling
cell migration. The traditional way to apply active gel theory to cell migration is to assume
that the polymerization at the cell membranes provides kinematic boundary conditions  
\cite{Kruse_2005_EPJE_Theory_Active_Polar_Gels_Paradigm_Cytoskeleton,
Juelicher_2007_PhyRep_Review_Active_Behavior_Cytoskeleton,
Kruse_2018_LectureNotes_Hydrodynamics_Active_Cytoskeleton}. One disadvantage of
this approach is that cell length then follows as a dynamical variable that cannot be controlled by
other means. In general, this approach is well suited to explain 
cell migration optogenetically controlled through actin polymerization. Here, however, we
are interested in the complementary situation that cell migration is controlled by contractility. 
This question has been addressed before in the framework of active gel theory by
using an elastic boundary condition \cite{Putelat_2018_PRE_stress_regulator_cell_motility, recho_contraction-driven_2013,
Recho_2015_JMPS_Mechanics_motility_initiation_arrest}, which implies that
cell size is controlled by elasticity-related processes, representing the 
outcome of the interplay between tension in the cell contour, bulk compressibility of the cell,
control of water flux through the membrane and adhesion to the environment
\cite{DizMunoz_2013_CellBio_Membrane_Tension_organizer_cell_shape_motility, 
Barnhart_2011_PLOS_Adhesive_dependent_mechanisms_motile_cell_shape}. Using an active gel model which includes
the myosin II concentration field, it has been shown that motile solutions are
possible for intermediate levels of contractility \cite{recho_contraction-driven_2013,
Recho_2015_JMPS_Mechanics_motility_initiation_arrest}. Here we follow a similar route, but
use an even simpler version of the elastic boundary model, which does not take
myosin II concentration into account. In the spirit of a minimal model approach,
we also disregard the effect of adhesion sites, which are known to lead
to nonlinear processes (in particular stick-slip oscillations) in the context of
cell migration \cite{sabass2010modeling,Ziebert_2013_PLOS1_Adhesion_substrate_shape_motility_crawling,sens_stickslip_2020,ron_one_2020}. Using the minimal active gel model with elastic boundary conditions, we can perform a comprehensive mathematical analysis of optogenetic control as a spatiotemporally constrained perturbation to the active stress. To make better contact to the situation in cells, we finally extend our analysis to boundary conditions with protrusion.

This work is organized as follows.
We start by introducing the minimal model for intracellular flows.
We consider a one-dimensional Maxwell model with active stresses that drive intracellular flow
with a frictional coupling to the environment. We show that such a model can
be induced to migrate due to optogenetic activation, but that this migration will stop when
optogenetic activation is turned off due to re-symmetrization of intracellular flow. We also show
that this model is able to qualitatively describe the dynamics of the actomyosin network 
along one-dimensional stress fibers during optogenetic activation. On this background, we then study the effect of 
including polymerization at the boundaries. 
We find that only asymmetric
situations can lead to the emergence of a motile state. We close with a summary of our results and an outlook on potential further research.

\section{\label{sec:Active_Maxwell}Minimal active gel model}

\subsection{Model definition}

To describe the effects of optogenetic perturbations, we consider a one-dimensional (1D) section across the cell. 
A minimal approach has to account for the following facts:
(i) the cytoplasm is mainly viscous, i.e.~it will flow at long time scales, while it 
can sustain stresses at short time scales. We hence use a Maxwell model,
where we can consider the purely viscous limit if needed for further simplification.
(ii) Flows inside the cell are balanced by friction forces with the substrate the cell sits on.
For simplicity we neglect focal adhesions and inhomogeneities and assume a homogeneous friction coefficient.
(iii) Concerning the 1D boundary conditions, one has to account for the fact that cells tend
to keep a typical size which is the result of the interplay between different
processes, including cortical tension and bulk compressibility. We 
hence implement an effective spring with a certain rest length and stiffness.
(iv) The cell is an active material, continuously converting 
metabolic energy into local motion, hence an active contractile stress is considered in the bulk.
The overall activity level is assumed to be homogeneous across the cell, while
optogenetic perturbations can be modeled as local, spatiotemporal changes in this contractility.

\begin{figure}[]		
 \centering
 \includegraphics[width=\columnwidth]{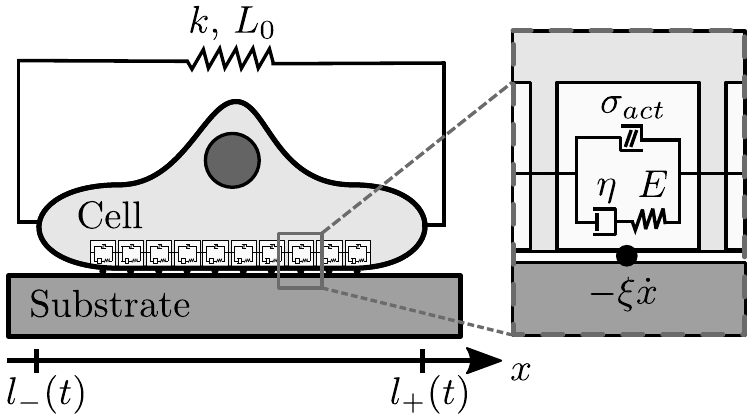}
 \caption[Scheme of the active Maxwell model with elastic boundaries]
 {Scheme of the active Maxwell model. The cell has a variable length $L=l_+-l_-$
 where $l_\pm$ denote its boundaries in one dimension. The cell's interior is described by the rheological model 
 of a spring with elastic modulus $E$ coupled in series to a dashpot with viscosity $\eta$, 
 comprising a Maxwell element. In addition, we demand the material to be frame-invariant, leading to properties of a co-rotational Maxwell material. This element is coupled in parallel to the active stress $\sigma_{act}$. 
 To the surrounding, the coupling is via viscous friction with friction coefficient $\xi$. 
 We apply elastic boundary conditions (stiffness $k$, rest length $L_0$), 
 and consider the continuum limit.}
 \label{fig:Active_Maxwell_rheological_model_schematic}	
 \end{figure}

Fig.~\ref{fig:Active_Maxwell_rheological_model_schematic} shows a schematic sketch of our model. As it is common in active gel theory, the active stress $\sigma_{act}$ is coupled in parallel to an infinitely compressible (co-rotational) Maxwell element describing the passive response of the cell.
Assuming that there is only a velocity in $x$-direction, the constitutive relation relating the total stress $\sigma$ and the strain $\epsilon$ 
is then \cite{Kruse_2005_EPJE_Theory_Active_Polar_Gels_Paradigm_Cytoskeleton, Juelicher_2007_PhyRep_Review_Active_Behavior_Cytoskeleton, Recho_2013_PRE_Pushing_pulling_crawling}
\begin{equation}\label{eq:Active_Maxwell_full_constitutive_eq_Maxwell}
\eta\partial_t{\epsilon}= [1+\tau\partial_t+\tau v \partial_x] \left(\sigma -\sigma_{act}\right),
\end{equation}
with $\tau=\eta/E$ defining the Maxwell relaxation time,
$E$ and $\eta$ being the elastic (shear) modulus and the viscosity, respectively, and $v=\dot{x}$ the flow velocity.
The cell is coupled to the substrate via viscous friction
\begin{equation}\label{eq:Active_Maxwell_viscous_friction}
\partial_x \sigma = \xi v,
\end{equation}
with $\xi$ being the homogeneous friction coefficient. 
Using $\partial_t{\epsilon}=\partial_x v$ yields an equation for the stress only 
\begin{equation}\label{eq:Active_Maxwell_governing_eq_stress}
\frac{\eta}{\xi} \partial_x^2 \sigma = \left[1+\tau \partial_t + \frac{\tau}{\xi}(\partial_x \sigma) \partial_x\right] \left(\sigma -\sigma_{act}\right).
\end{equation}
The length scale $\sqrt{\eta/\xi}$
is known as the hydrodynamic decay length. In the following
we will assume that the active background stress $\sigma_{act}$
is a constant, so the terms $\partial_t \sigma_{act}$ and $\partial_x \sigma_{act}$ will 
disappear. However, they will reappear for
optogenetic activation due to its dependence on time and space.

As a spreading or moving cell is a moving boundary problem, one has to consider the 
left and right boundary, $l_-(t)$ and  $l_+(t)$, of the cell 
and consequently the cell's length $L(t)=l_+(t)-l_-(t)$,
as functions of time. 
We consider an elastic boundary condition
\cite{Putelat_2018_PRE_stress_regulator_cell_motility, recho_contraction-driven_2013,Recho_2015_JMPS_Mechanics_motility_initiation_arrest}
\begin{equation}\label{eq:Active_Maxwell_BC_stress}
  \sigma(l_\pm(t),t) = - k \frac{L(t)-L_0}{L_0},
  \end{equation}       
with reference length $L_0$ and effective spring constant $k$. 
The  boundaries are assumed to flow with the gel, i.e.~the velocity $v$ 
there is also given by Eq.~(\ref{eq:Active_Maxwell_viscous_friction}),
  \begin{equation}\label{eq:Active_Maxwell_BC_dot_lpm}
  \dot{l}_\pm(t) = v(l_\pm(t)) = \frac{1}{\xi}\partial_x \sigma(x=l_\pm(t),t).
  \end{equation}
We non-dimensionalize the equations by rescaling length by $L_0$, 
time by $\xi L_0^2/k$ and stress by $k$ 
to obtain the boundary value problem (BVP)
\begin{equation}\label{eq:Active_Maxwell_nondim_bvp}
  \begin{gathered}
  \mathcal{L}^2 \partial_{x}^2 \sigma - \mathcal{T} \partial_t \sigma - \mathcal{T}(\partial_x \sigma)^2 - \sigma = -\sigma_{act}, \\
  \sigma(l_\pm(t),t) = -(L(t)-1), \\
  \dot{l}_\pm = \partial_x \sigma(l_\pm(t),t),
  \end{gathered}
\end{equation}
with only two dimensionless parameters:
a relative length scale $\mathcal{L}=\sqrt{\eta/(\xi L_0^2)}$ comparing viscous and frictional damping of the cytoplasm flow 
and a renormalized Maxwell relaxation time $\mathcal{T}=(k\tau)/(\xi L_0^2)=(k\eta)/(\xi E L_0^2)$. 
Note that all mechanical/dynamical parameters, $\eta$, $E$, $k$, $\xi$, together with the rest length, 
determine the relaxation time of the system as a whole.  We also note that the purely viscous case
corresponds to $\mathcal{T}=0$; in this case two terms disappear from Eq.~(\ref{eq:Active_Maxwell_nondim_bvp}), 
a linear one often considered as an approximation for small flow velocities in the so-called linear Maxwell model, 
and a nonlinear one resulting from the frame invariance.
  
\subsection{\label{sec:Active_Maxwell_steady_states}Steady state solutions}
\label{stst}

We first determine the possible steady state solutions
by assuming a constant cell velocity $V$ and no length change, $\dot{L}=0$. 
It is useful to map the problem on the unit interval 
by changing into internal coordinates $u=(x-l_-)/L$.
To simplify the boundary conditions we introduce the deviation of the stress from 
the elastic boundary contribution as
	\begin{equation}\label{eq:Active_Maxwell_definition_auxillary_s}
	s(u,t) = \sigma(u,t) + (L(t)-1).
	\end{equation}  
For a steady state solution the stress deviation $s$ can only depend 
on the internal coordinate $u$, which yields 
	\begin{equation}\label{eq:Active_Maxwell_nondim_bvp_s_u}
	\begin{gathered}
		 \frac{\mathcal{L}^2}{L^2}\partial_{u}^2 s + \mathcal{T}\frac{V}{L}\partial_u s - \frac{\mathcal{T}}{L^2} (\partial_u s)^2 - s + (L-1) = -\sigma_{act}, \\
		s(u_\pm) = 0, \qquad \partial_u s(u_\pm) = VL
	\end{gathered}
	\end{equation}
with $u_-=0$ and $u_+=1$.

This equation can be rewritten as a dynamical system, i.e.~two first order differential equations as given in appendix \ref{appendix_lack_motile_steadstate}. For non-motile solutions, i.e.~$V=0$, the boundary conditions then imply that the solution corresponds to a fixed point with $s\equiv 0$ and
a condition for the length of the non-motile steady state solution,
\begin{equation}\label{eq:Active_Maxwell_stable_Lhat}
 \hat{L}=1-\sigma_{act}.
\end{equation} 
Hence the length of the cell decreases with increasing contractile active stress.
This solution ceases to exist for too large stresses, $\sigma_{act}>1$, 
when the elastic boundary condition cannot counteract the contractile stress anymore. 

Concerning motile solutions with $V\neq0$, it can be shown that, due to the boundary conditions, such solutions must correspond to periodic orbits in phase plane. However, the dynamical system is a gradient system with a potential, given by Eq.~(\ref{eq:Lack_motile_steady_state_potential}), and thus no periodic orbits exist \cite{Strogatz_2015_Book_Nonlinear_dynamics}.
Note that this is also true for the approximation of a linear Maxwell material, neglecting the quadratic term $(\partial_x\sigma)^2$ in Eq.~(\ref{eq:Active_Maxwell_nondim_bvp}), which can be considered as an approximation for small flow velocities.
Details are given in appendix \ref{appendix_lack_motile_steadstate}.

\subsection{\label{sec:Active_Maxwell_analytical_viscous}Analytical solution for the purely viscous case}

For the internal dynamics of cells, often it is assumed that the elastic component can be neglected, 
since the relaxation time is much smaller than the 
experimental timescales observed for processes like cell spreading or motility
\cite{Rubinstein_2009_BioPhysJ_Viscpelastic_flow_keratocyte_lemellipod}. 
The purely viscous case with $\mathcal{T}=0$ can be solved analytically: 
one can integrate Eq.~(\ref{eq:Active_Maxwell_nondim_bvp_s_u}) directly 
and find the Green's function, allowing to obtain
the general solution for arbitrary lengths $L$
\begin{equation}\label{eq:Active_Maxwell_solution_s_viscous}
  s(u,t) = \Big(\sigma_{act} + (L-1)\Big) \left( 1- \frac{\cosh\left(\frac{L}{\mathcal{L}}(u-1/2)\right) }{\cosh(L/2\mathcal{L})}\right).
\end{equation}
Using this solution and Eq.~(\ref{eq:Active_Maxwell_nondim_bvp}), 
one gets the governing equation for the cell length, 
$\dot{L}=\dot{l}_+-\dot{l}_-=(\partial_us(1)-\partial_u s(0))/L$, 
\begin{equation}\label{eq:Active_Maxwell_ode_for_length_viscous}
\dot{L}=\frac{2}{\mathcal{L}} \big( 1-L-\sigma_{act} \big)  \tanh(L/2\mathcal{L}).
\end{equation}

Since $L>0$
one has $\dot{L}<0$ ($>0$) if $L>1-\sigma_{act}$ ($<1-\sigma_{act}$) 
and hence a relaxation towards the non-motile steady state. 

For small perturbations around the steady state $\hat{L}=1-\sigma_{act}$ 
we expand the length as $L(t)=\hat{L}+\delta L(t)$ (with $\delta L(t=0)= \delta L_0$) up to second order and get
\begin{equation}\label{eq:Active_Maxwell_second_order_perturbative_solution_L}
\delta L(t) = \frac{\alpha\ \delta L_0\ \exp(\alpha t)}{\alpha - \beta\ \delta L_0\ (\exp(\alpha t) -1)}
\end{equation}
with 
\begin{equation}
\alpha =-\tfrac{2}{\mathcal{L}}\tanh\left(\hat{L}/2\mathcal{L}\right),\,\,
 \beta = -\tfrac{1}{\mathcal{L}^2}\left(1-\tanh^2\left(\hat{L}/2\mathcal{L}\right)\right).\nonumber
\end{equation}
This solution is an exponential relaxation towards the stationary length $\hat{L}$ with 
relaxation time $1/\alpha$ and a higher order correction.

\subsection{\label{sec:Active_Maxwell_numerical_nooptgen}Numerical solution}

To numerically solve our general model, we slightly reformulated the boundary value problem,
introducing the rescaled stress $\hat{\sigma}=L\sigma$ and the cell's center position $G=(l_++l_-)/2$.
This procedure results in an advection-diffusion
equation with source term that can be solved with the Finite Volume Method. 
Details can be found in appendix \ref{appendix_numerics}.

\begin{figure}[]
\centering
 \includegraphics[width=\columnwidth]{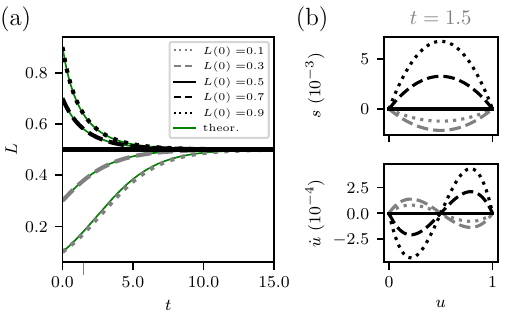}
\caption[Numerical results for the active Maxwell model with elastic boundary 
conditions and varying $L(0)$]{
Finite Volume simulations for the purely viscous model. 
Part (a) shows the resulting cell length $L$ (in units of the rest length $L_0$) 
as a function of time (numerical/analytical results in black/green).
Part (b) shows the stress deviation from the boundary condition $s=[\sigma+(L-1)]$ 
and the internal velocity field $\dot{u}$, 
for the time point indicated by the grey axis tick in (a). 
Parameters: $\mathcal{T}=0$, $\mathcal{L}=1$, $\sigma_{act}=0.5$.}
\label{fig:Active_Maxwell_basemodel_simulation_varying_L_starting}
\end{figure}

We first studied the viscous case,  $\mathcal{T}=0$, as shown 
in Fig.~\ref{fig:Active_Maxwell_basemodel_simulation_varying_L_starting}. 
Starting with a stress-free slab ($s\equiv0$) and $\dot{L}=\dot{G}=G=0$, we see a relaxation behavior of the cell's length
towards the stationary solution with $L=\hat{L}$ for different starting lengths $L(0)$,
cf.~Fig.~\ref{fig:Active_Maxwell_basemodel_simulation_varying_L_starting}(a). 
The numerical result is in good agreement with the second order perturbative 
solution given in Eq.~(\ref{eq:Active_Maxwell_second_order_perturbative_solution_L}), cf.~the green curves. 
In the cell's interior, there is the typical $\cosh$-shaped stress profile
which is contractile or extensile, as well as anterograde or retrograde flow 
inside the cell, see Fig.~\ref{fig:Active_Maxwell_basemodel_simulation_varying_L_starting}(b),
again 
in agreement with the analytical solution Eq.~(\ref{eq:Active_Maxwell_solution_s_viscous}). 
Note that the flow inside the cell is negligible 
and that 
the length change dominates the material flow in the lab frame. 
Our simulations also show that increasing (decreasing) $\mathcal{L}$ 
leads to slower (faster) relaxation, which can be traced back to $\mathcal{L}^2/L^2$ 
being the effective diffusion constant (see Eq.~(\ref{eq:Active_Maxwell_numerical_convection_advection})). 
Changes in $\mathcal{L}$, however, do not change the results qualitatively.
Changing the active stress $\sigma_{act}$ only shifts the stable length $\hat{L}$ 
according to Eq.~(\ref{eq:Active_Maxwell_stable_Lhat}), as long as $\sigma_{act}<1$.

\begin{figure}[!t]
	\centering
	\includegraphics[width=\columnwidth]{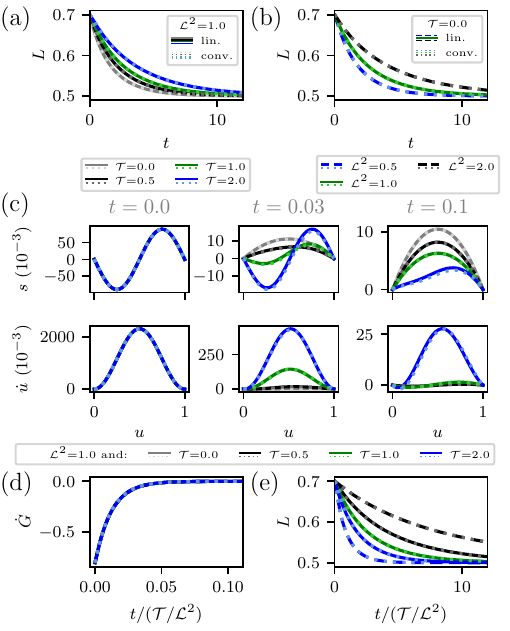}
	\caption[Numerical results for the active Maxwell model with elastic boundary conditions and 
	biased initial conditions]{
	Finite Volume simulations for the active Maxwell model starting with an asymmetric initial stress,
	showing re-symmetrization and motility arrest.
	Panel (a) and (b) show length relaxations ($L$ in units of $L_0$) for $\mathcal{L}^2=1$ and varying $\mathcal{T}$,  and for $\mathcal{T}=0$ and varying $\mathcal{L}^2$, respectively. 
	The stress deviations from the boundary condition $s$
	and the internal velocity fields $\dot{u}$ 
	are shown in panel (c) for the stated time points for fixed $\mathcal{L}^2=1$.
	Panels (d) and (e) show the cell velocities $\dot{G}$ and cell lengths $L$ of all simulations in (a-c), respectively, with time rescaled by $\mathcal{T}/\mathcal{L}^2$. 
	In (d) all curves collapse onto a single curve, showing that the re-symmetrization timescale is determined by $\mathcal{T}/\mathcal{L}^2$. 
	The solid/dashed lines correspond to the linear and the dotted ones to the (co-rotational) convected Maxwell model. 
	The models show only slight deviations during re-symmetrization.  Parameters: $\sigma_{act}=0.5$, corresponding to $\hat{L} = 0.5$.}
\label{fig:Active_Maxwell_basemodel_simulation_biased_initials}
\end{figure}	

To understand the influence of the viscoelastic material properties, 
we next considered different Maxwell relaxation times $\mathcal {T}$,
but now starting with a symmetry-broken initial condition for the stress
for which we by way of example chose
\begin{equation}
\hat{\sigma}(u,0)=-L(0)(L(0)-1)\ [1+0.3 \sin(2\pi u)].
\end{equation}
The results are summarized in 
Fig.~\ref{fig:Active_Maxwell_basemodel_simulation_biased_initials}.
Fig.~\ref{fig:Active_Maxwell_basemodel_simulation_biased_initials}(a) shows
that increasing the Maxwell time leads to slower length relaxations.
Fig.~\ref{fig:Active_Maxwell_basemodel_simulation_biased_initials}(b) exemplifies our prior observation that, for the purely viscous case, increasing the viscous length scale leads to slower relaxation.
As expected from the lack of motile steady states discussed above, 
the symmetry-broken initial state is rapidly resymmetrized in all cases,
cf.~Fig.~\ref{fig:Active_Maxwell_basemodel_simulation_biased_initials}(c), which shows the time development by plots for three different time points.
Interestingly, this re-symmetrization happens on a much faster time scale than 
the one for the length relaxation process. 
Note that for the purely viscous model ($\mathcal{T}=0$), re-symmetrization is instantaneous, 
as a consequence of the symmetrical solution, Eq.~(\ref{eq:Active_Maxwell_solution_s_viscous}),  only depending on $L$. 
The broken symmetry of the initial state induces a transient net cell velocity $\dot{G}$, which goes to zero as the cell resymmetrizes, 
cf. Fig.~\ref{fig:Active_Maxwell_basemodel_simulation_biased_initials}(d).  Rescaling time by $\mathcal{T}/\mathcal{L}^2$ in simulations for different $\mathcal{T}$ and $\mathcal{L}^2$ in (d,e)
shows that all cell velocities $\dot{G}$ collapse onto a single curve, implying that $\mathcal{T}/\mathcal{L}^2$ determines the time scale of re-symmetrization.
Fig.~\ref{fig:Active_Maxwell_basemodel_simulation_biased_initials}(e) shows that this is not true for the length relaxation, which agrees to the nonlinear dependence of the relaxation timescale on $\mathcal{L}^2$ of the perturbative viscous ($\mathcal{T}=0$) solution Eq.~(\ref{eq:Active_Maxwell_second_order_perturbative_solution_L}). Neglecting the term quadratic in $\partial_x\sigma$ in Eq.~(\ref{eq:Active_Maxwell_nondim_bvp}) corresponds to the linear Maxwell model, which constitutes a small flow velocity approximation. This simplification yields similar results with only small deviations during re-symmetrization, cf.~Fig.~\ref{fig:Active_Maxwell_basemodel_simulation_biased_initials}(c), in agreement to the small internal flow velocities $\dot{u}$. Note that these differences are so small that they do not visibly change the length and velocity dynamics, cf.~Fig.~\ref{fig:Active_Maxwell_basemodel_simulation_biased_initials}(a,b,d).

At this point we can conclude that the model gives a good description of 
how the actomyosin system flows inside an adherent cell with a typical size.
In the next section we therefore can turn to the effect of
optogenetic perturbations. As shown in Fig.~\ref{fig:Active_Maxwell_basemodel_simulation_biased_initials}, 
more complex material properties (namely visco\-elastic rather than viscous) allow the system to have 
short periods of motility, but as the asymmetry relaxes rapidly, also the movement rapidly stops.
The model is hence unable to describe self-polarization due to the lack of 
motile states as found in section \ref{stst}.

\section{\label{sec:Active_Maxwell_optgen}Optogenetic control}

\subsection{Model definition}

We now turn to the effect of optogenetic control of the cytoskeleton. 
Different experimental strategies have been implemented, but most of them
are similar in the sense that one engineers a light-sensitive process into
the cellular control circuits for the cytoskeleton. Typically the effect
of light is to recruit a GTP-exchange factor to the membrane, where it
activates a member of the Rho-family of small GTPases, e.g.\ Rac1 for
actin protrusions \cite{wu_genetically_2009,wang_light-mediated_2010,valon_predictive_2015,Kato_2015_PLOS1_Lamellipodial_extension_optogenetics} or RhoA for actomyosin contraction \cite{Oakes_2017_NatComm_Optogenetic_RhoA, Valon_2017_Nature_Optogenetic_control_cellular_forces}. In the latter case,
which is of special interest for the present work, the main effect is
a local increase in active stress $\sigma_{act}$ due to actin polymerization and assembly of
myosin II minifilaments.

The effect of a localized optogenetic activation can be incorporated in the model 
by adding a spatiotemporally constrained additional term to the active stress. Experimentally the used laser spots have typical
spatial profiles that can easily be described mathematically.
The resulting biochemical activation has a kinetic 
profile determined by reactions and diffusion that can be measured e.g. by using fluorescent 
probes \cite{valon_predictive_2015,kamps_optogenetic_2020}.
To describe these processes, here we introduce a dimensionless spatiotemporal ``shape function'' 
 $\Xi(x,t)$
and replace $\sigma_{act}\rightarrow\sigma_{act}+\sigma_{opt}$, where $\sigma_{act}$ is, as before,
the homogeneous contractile stress and $\sigma_{opt}=\varepsilon\,\Xi(x,t)$ is the localized
contribution from optogenetic activation.

Rescaling the optogenetic stress level $\varepsilon$ with $k$, 
we obtain instead of Eq.~(\ref{eq:Active_Maxwell_nondim_bvp})
the modified non-dimensional BVP
  \begin{multline}\label{eq:Active_Maxwell_optgen_nondim_bvp}
  \mathcal{L}^2 \partial_{x}^2 \sigma - \mathcal{T} \partial_t \sigma - \mathcal{T}(\partial_x \sigma)^2 - \sigma 
  = \\ -\sigma_{act} - \varepsilon\,\Xi - \varepsilon \mathcal{T} \partial_t \Xi - \varepsilon \mathcal{T}(\partial_x\sigma)(\partial_x\Xi),
  \end{multline}
where the boundary conditions and equations determining the position and length are unchanged. Note the existence of the last two terms, stemming from the time- and space-dependence of the shape function.
  
As suggested by the experimental situation, we assume that $\Xi$ factorizes,
\begin{equation}
  \Xi(x,t) =   \Xi_t(t) \cdot \Xi_s(x,t),
\end{equation}
with one factor describing 
the turn on and turn off process of the signal, which only depends on time $t$, 
and one factor describing the spatial shape of the signal. 
Importantly, the latter depends on the spatial coordinate $x$ but also on $t$, 
due to changes in the cell's length (reflected by the internal coordinate $u$).

We consider two main protocols: either the signal is considered to be fixed in the stationary lab coordinates
or it is considered to be fixed in co-moving coordinates 
(i.e.~it is moved along with the cell). 
The former is what is typically -- for simplicity -- realized experimentally,
but the latter should also be realizable
and is instructive concerning cell motility. 

From our previous analysis we know that the model 
has only the non-motile steady state given by Eq.~(\ref{eq:Active_Maxwell_stable_Lhat}).
After the signal has been turned off again, i.e.~for $\Xi_t(t) \approx 0$, 
we have the same equations studied in the last section, implying
relaxation to the only stable solution, provided the length remains in the stable regime $L>0$. 
The optogenetic signal should therefore only 
induce a perturbation of the known steady state during activation.

To get a first insight into the effect of the perturbation,
we consider a steady state solution (with $\dot{L}=0$, $\dot{G}=V$, with $V=0$ describing the resting state) 
for the purely viscous case ($\mathcal{T}=0$)
and a co-moving stress field and shape function, 
i.e.~$\sigma=\sigma(u)$ and $\Xi=\Xi(u)$. 
Using again the stress deviation $s$, in analogy to Eq.~(\ref{eq:Active_Maxwell_nondim_bvp_s_u}) we now have
  \begin{equation}\label{eq:Active_Maxwell_optgen_nondim_bvp_s_u}
		 \mathcal{L}^2 \frac{1}{L^2}\partial_{u}^2 s - s  = - (L-1) - \sigma_{act}  - \varepsilon \Xi,
  \end{equation}
with the same boundary conditions $s(u_\pm) = 0$ and $\partial_u s(u_\pm) = VL$.
This equation can be integrated 
to obtain an equation for the perturbed  length 
\begin{equation}\label{eq:Active_Maxwell_optgen_full_equation_hat_L}
\hat{L}=1-\sigma_{act}-\varepsilon\int_0^1 \Xi(u) \mathrm{d}u+\int_0^1 s(u) \mathrm{d}u.
\end{equation}

The stationary length is therefore changed in the following way:
the integrated signal gives an additional active stress, which tends to decrease 
the length of the cell, while the resulting stress profile $s$ in the steady state counteracts 
this active stress, since the active stress is not compensated by the boundary stress from cell contraction.
Therefore a total contractile stress in the non-motile steady state, $\int s \mathrm{d}u>0$, 
leads in principle to a larger length. However, we will see in the numerical simulations 
that this contribution from the stress field is negligible. 
Therefore the integral over the optogenetic signal stress can be used as a good estimate 
to determine the perturbed 
stationary length.

For the numerical simulations
we use the auxiliary stress field in internal coordinates
$\hat{\sigma}(u)=L\sigma(l_-+uL,t)$.
The two protocols, i.e.~the signal either fixed 
in co-moving or in lab coordinates, are denoted with superscripts 
cm and x, respectively.

For the shape function in internal coordinates, 
we mainly use a box-shaped spatial shape function $\Xi_s$ 
that spans from $u_\mathrm{on}$ to $u_\mathrm{off}$:  i.e.~
$\Xi_\mathrm{box} (u,0)=1$ for $u\in [u_\mathrm{on},u_\mathrm{off}]$ and 
$\Xi_\mathrm{box} (u,0)=0$ otherwise.
To test the influence of the details of the spatial shape of the activation spot,
we also considered a smooth Gaussian shape function (with the center 
located at $(u_\mathrm{on}+u_\mathrm{off})/2$, standard deviation  
$\omega=(u_\mathrm{on}-u_\mathrm{off})/2$ 
and normalization 
such that the integral
is identical to the one of the box-shaped spatial shape function). 

For the temporal switching behavior (temporal shape function) we also considered several signal types: 
introducing a turn on time $t_\mathrm{on}$ and a turn off time $t_\mathrm{off}$, 
we use a box-shaped temporal shape function $\Xi_t^\mathrm{box}$, 
a continuous $\tanh$-type box function $\Xi_t^\mathrm{tanh}$ and 
an exponentially plateauing activation function $\Xi_t^\mathrm{exp}$. 
Note that the latter has been used successfully  to describe the activation behavior and reaction time
of signaling pathways in optogenetic experiments on stress fiber dynamics \cite{Oakes_2017_NatComm_Optogenetic_RhoA}.
Specific formulas for all shape functions are given in appendix \ref{app:Shape_Functions}.

\subsection{\label{sec:Active_Maxwell_optgen_centered}Centered activation}

We consider the purely viscous model.
We solved it
numerically, using the stationary solution with homogeneous stress as initial data, 
to study the response to an optogenetic activation in the cell's center. 
\begin{figure}[!t]
	\centering
		\includegraphics[width=\columnwidth]{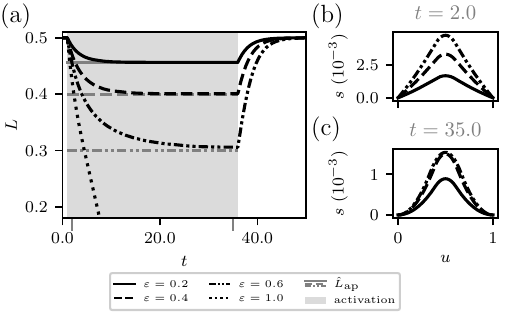}
	\caption[Numerical results for the active Maxwell model with elastic boundary conditions and varying $L(0)$
	for centered activation]
	{Finite Volume simulations for the viscous model with a cell-centered activation 
	(with box-shaped temporal and spatial shape functions) 
	for different activation strengths $\varepsilon$. 
	Shown in (a) is the resulting length of the cell $L$ (in units of $L_0$) 
	as a function of time for a signal that is fixed 
	in co-moving coordinates.
	The shaded area shows the activation time from $t_\mathrm{on}=1$ to $t_\mathrm{off}=36$ 
	and the horizontal lines indicate the approximate theoretical new stationary length $\hat{L}_\mathrm{ap}$, 
	cf.~Eq.~(\ref{eq:Active_Maxwell_optgen_approx_equation_hat_L}). 
	Parts (b) and (c) show the stress deviation from the boundary condition $s=[\sigma+(L-1)]$ during early and late 
	activation, respectively, with timepoints indicated as grey axis ticks in (a) and given above. 
	Parameters: $\mathcal{T}=0$, $\mathcal{L}=1$, $\sigma_{act}=0.5$, $u_\mathrm{on}=0.4$, $u_\mathrm{off}=0.6$.
	}
\label{fig:Active_Maxwell_optgen_simulation_comparison_epsilons}
\end{figure}
Fig.~\ref{fig:Active_Maxwell_optgen_simulation_comparison_epsilons}
shows results for an optogenetic signal applied in the co-moving frame 
(as the cell does not move, lab frame yields the same result)
for varying strength of the optogenetic signal $\varepsilon$. 
The signal was chosen to be box-shaped in both space and time. %
One clearly sees that the cell length relaxes to a new stable length while the signal is turned on. 
The signal strength $\varepsilon$ determines how much the cell is contracted. 
This contraction can be so large,  
that for $\varepsilon=1$, $L\rightarrow 0$,
cf.~Eq.~(\ref{eq:Active_Maxwell_stable_Lhat}).
Fig.~\ref{fig:Active_Maxwell_optgen_simulation_comparison_epsilons} 
also shows as horizontal lines the approximate theoretical result, cf.~Eq.~(\ref{eq:Active_Maxwell_optgen_full_equation_hat_L}),
which in the co-moving frame reads
  \begin{equation}\label{eq:Active_Maxwell_optgen_approx_equation_hat_L}
  \hat{L}_\mathrm{ap}=1-\sigma_{act} -\varepsilon \int_0^1  \Xi(u; \hat{L}_\mathrm{ap})\mathrm{d}u.
  \end{equation}
Hence in this case 
the integral over $\Xi$ 
in internal coordinates depends on the length $\hat{L}_\mathrm{ap}$, yielding a quadratic equation in the length. 
  The obtained theoretical length predicts the new stationary length 
  for the activated cell
  very well, 
  suggesting that the integral of the activation term 
  is indeed the determining quantity for the length.

\begin{figure}[!t]
	\centering
		\includegraphics[width=\columnwidth]{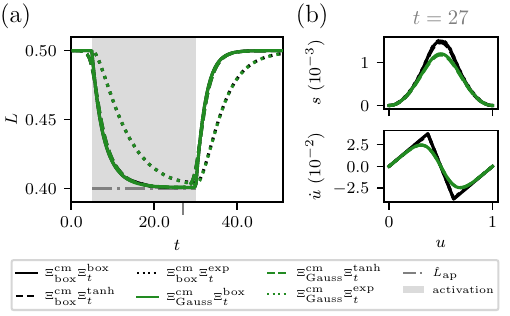}
	\caption[Numerical results for the active Maxwell model with elastic boundary conditions and varying $L(0)$ 
	for centered activation]{
	Finite Volume simulations for the viscous model for centered activation in cm frame 
	with different shape functions. 
	Part (a) shows the resulting length of the cell $L$ (in units of $L_0$) as a function of time 
	and the theoretical approximation for the stationary perturbed length $\hat{L}_\mathrm{ap}$ 
	from (\ref{eq:Active_Maxwell_optgen_approx_equation_hat_L}).
	Part (b) shows the stress deviation from the boundary condition $s=[\sigma+(L-1)]$ and the 
	internal velocity field $\dot{u}$ for the time points indicated by the grey axis tick in (a) and given above. 
	Different optogenetic signals $\Xi$ where applied as specified in the legend.
	Parameters: $\mathcal{T}=0$, $\mathcal{L}=1$, $\sigma_{act}=0.5$, $\varepsilon=0.4$, $u_\mathrm{on}(t=0)=0.4$, $u_\mathrm{off}(t=0)=0.6$.
	}
\label{fig:Active_Maxwell_optgen_simulation_comparison_time_ramp}
\end{figure}

Fig.~\ref{fig:Active_Maxwell_optgen_simulation_comparison_time_ramp}
investigates
the effect of different shape functions for the optogenetic perturbation
for constant signal strength $\varepsilon=0.4$.
We used box vs.~Gaussian spatial shape functions
and the three temporal shape functions (box, tanh and exponential) 
modeling the turn on and turn off process. 
For Gaussian and box-shaped spatial shape functions 
the resulting new lengths coincide, which again
shows the dependence on the integrated signal only, cf.~Eqs.~(\ref{eq:Active_Maxwell_optgen_full_equation_hat_L},\ref{eq:Active_Maxwell_optgen_approx_equation_hat_L}). 
Also, both signals lead to similar stress profiles in the stationary activated state,
cf.~Fig.~\ref{fig:Active_Maxwell_optgen_simulation_comparison_time_ramp}(b), 
showing only 
small differences in the internal stress (deviation) field $s$ in the stationary perturbed state, 
but rather large ones in the internal flow field $\dot{u}$. 
There the fields with Gaussian activation are much smoother, due to the smoothness of the perturbation.

Concerning the temporal shape function $\Xi_t$ we used a box signal
(with $t_\mathrm{on}=6$ and $t_\mathrm{off}=31$),
 a smooth $\tanh$-box-function, see Eq.~(\ref{eq:appendix_shape_functions_t_tanh}),
 as well as an exponential plateauing function, Eq.~(\ref{eq:appendix_shape_functions_t_exp}). 
Naturally, smoother shape functions smoothen the dynamics. 
Nonetheless the same relaxation behavior towards the new stationary length after turning on the signal and towards 
the initial length after turning off the signal is obtained. 
The profiles in Fig.~\ref{fig:Active_Maxwell_optgen_simulation_comparison_time_ramp}(b) agree 
for the temporal shape functions in the region where the length is relaxed towards the new stationary length 
during optogenetic activation. 
Note that, for the given parameters, the exponential shape function yields slower dynamics 
and the new equilibrium length is  reached only at the end of activation.

These results suggest that the exact form 
of the temporal shape function 
is not of significance for the modeling of the cell's behavior, provided the turn-on and turn-off process 
is of short duration compared to the time of the optogenetic perturbation in total. 
For slowly changing activation the temporal shape function determines the exact shape 
of the length change (cf.~the exponential shape function). In the following we hence 
mostly focus on the box shape functions for simplicity.

\subsection{\label{sec:Active_Maxwell_optgen_asymm}Asymmetric activation and kymographs}

We know from our previous analysis that motile 
steady state solution 
states that are initially symmetry broken 
resymmetrize on a short timescale. 
We now investigate asymmetric optogenetic activation, actively breaking the symmetry in the cell, 
with the aim to effect motile states, at least when using co-moving activation.

We performed simulations with flat initial profiles to which
an optogenetic perturbation is applied with 
temporal and spatial box signals and 
an asymmetric spatial shape function $\Xi_s$. 
To roughly characterize the spatial asymmetry we introduce an asymmetry parameter
\begin{equation}\label{eq::modkruse_numerics_asymm_delta}
\delta = \frac{u_\mathrm{off}+u_\mathrm{on}}{2} - \frac{1}{2} \in \left[-\frac{1}{2},\ \frac{1}{2}\right],
\end{equation}  
where $\delta<0$ ($\delta >0$) characterizes the offset of the optogenetic signal 
to the left (right) with respect to the cell's center. 

\begin{figure}[!t]
	\centering
		\includegraphics[width=\columnwidth]{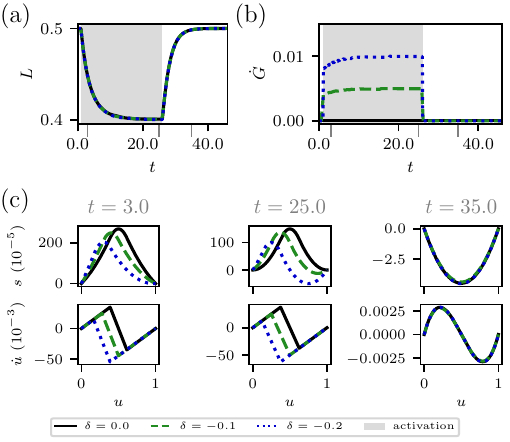}
	\caption[Numerical results for the active Maxwell model with elastic boundary conditions and varying $L(0)$ for centered activation]
	{Finite Volume simulations
	for asymmetric activation 
	with temporal and spatial box shape functions and for different asymmetry parameters $\delta$, cf.~Eq.~(\ref{eq::modkruse_numerics_asymm_delta}). 
	Part (a) shows the length of the cell $L$ (in units of $L_0$) as a function of time for a signal 
	fixed in co-moving coordinates. Part (b) shows the cell's velocity $\dot{G}$ 
	and in (c) the stress deviation from the boundary condition $s=[\sigma+(L-1)]$ 
	and the internal velocity field $\dot{u}$ are shown for the time points indicated 
	by the grey axis ticks in (a) and (b). The shaded area in (a), (b) marks the time window of activation, here 
	from $t_\mathrm{on}=1$ to $t_\mathrm{off}=26$. 
	Parameters: $\mathcal{T}=0$, $\mathcal{L}=1$, $L(0)=0.5$, $\sigma_{act}=0.5$, $\varepsilon=0.4$, $u_\mathrm{on}(t=0)=0.4+\delta$, $u_\mathrm{off}(t=0)=0.6+\delta$.
	}
\label{fig:Active_Maxwell_optgen_simulation_asymm_varying_delta}
\end{figure}

Fig.~\ref{fig:Active_Maxwell_optgen_simulation_asymm_varying_delta}
investigates the case of a co-moving activation signal.
Part (a) 
shows that the stationary length does not change with different asymmetry parameter $\delta$, 
which again verifies that the integrated optogenetic signal determines the dynamics of the cell's length. 
Part (b), however, shows that the cell velocity increases for signals with larger $\delta$, 
implying that optogenetic activation in the cell's periphery has a larger effect than in the cell's center. 
This is in accordance with the fact that a larger $\delta$ leads to larger asymmetry in the cell, 
as can be also seen in the profiles in Fig.~\ref{fig:Active_Maxwell_optgen_simulation_asymm_varying_delta}(c). 
During activation a larger asymmetry in the stress deviation and the flow profile can be observed. 
The velocity is positive for a negative $\delta$, meaning that asymmetric optogenetic signals 
that lead to additional contractions determine the polarity of the moving cell with the additional contraction 
in the trailing half, which is in accordance to both experimental \cite{Mogilner_2020_SemCellDevBio_Review_Keratocytes_Cell_Motility, Verkhovsky_1999_CurrBiol_Selfpolarization_directional_motility} and theoretical \cite{Callan_Jones_2013_NJP_active_gel_amoeboid_motility, Recho_2015_JMPS_Mechanics_motility_initiation_arrest}
observations. 
Note that we have re-symmetrization and motility arrest rapidly after the optogenetic 
signal has been switched off (at the end of the shaded grey area in
Fig.~\ref{fig:Active_Maxwell_optgen_simulation_asymm_varying_delta})
in accordance with our expectation from our previous results.

\begin{figure}[!t]
	\centering
	\includegraphics[width=\columnwidth]{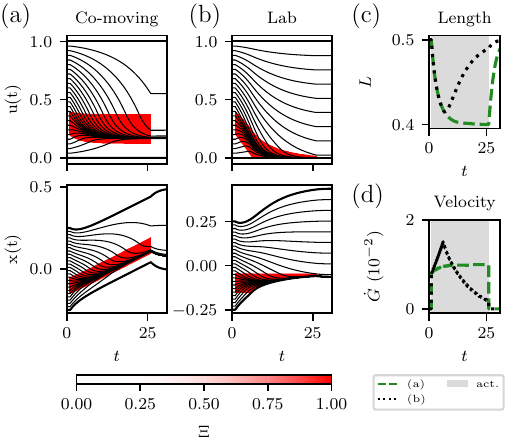}
	\caption{Shown are kymographs (space-time plots) for an applied optogenetic activation
	signal that is co-moving  with the cell (a) or fixed in the lab frame (b), respectively. 
	The curves traced in (a) and (b) correspond to trajectories of material points and the value of the box shape function $\Xi$, 
	corresponding to the optogenetic activation strength, is depicted in red. The thick lines are the cell boundaries. 
	The upper panels in (a) and (b) show the flow in internal coordinates ($u$) and 
	the lower panels in the lab frame ($x$). 
	Panels (c) and (d) show the resulting cell length and velocity, respectively.
	Asymmetric box signals with $u_\mathrm{on}(t=0)=0.2, u_\mathrm{off}(t=0)=0.4$, $t_\mathrm{on}=1, t_\mathrm{off}=26$ 
	were applied. 
	Other parameters:  $\mathcal{T}=0$, $\mathcal{L}=1$, $\sigma_{act}=0.5$, $\varepsilon=0.4$.}
\label{fig:Active_Maxwell_optgen_simulation_asymm_kymograph_coordsys}
\end{figure}

We next compared the two different experimentally accessible protocols -- fixing the activation in either the
co-moving or the lab frame -- again using box-shaped signals.
Fig.~\ref{fig:Active_Maxwell_optgen_simulation_asymm_kymograph_coordsys} shows 
kymographs, i.e.~the positions of material points as a function of time, and the cell length and velocity as previously. 
The kymographs are plotted twice, the upper panels using the internal coordinate system ($u$)
and the lower panels the fixed lab coordinates ($x$). 

Fig.~\ref{fig:Active_Maxwell_optgen_simulation_asymm_kymograph_coordsys}(a) 
shows the result for a perturbation
fixed in the co-moving frame.
We see that the signal in this frame moves with the cell in the lab system (lower panel), 
but stays fixed in width upon length changes of the cell, as it should. 
After a short adaptation period, the cell's velocity becomes constant and the cell attains a constant velocity, 
as can be also seen from Fig.~\ref{fig:Active_Maxwell_optgen_simulation_asymm_kymograph_coordsys}(c),(d). 
During the activation the material flows towards the activation center, which can also be seen 
in the flow profiles $\dot{u}$ in Fig.~\ref{fig:Active_Maxwell_optgen_simulation_asymm_varying_delta}(c).
Finally, as expected, the motility arrests shortly after the signal ends. 

Fig.~\ref{fig:Active_Maxwell_optgen_simulation_asymm_kymograph_coordsys}(b) shows the simplest experimental protocol, 
where the signal is fixed in the lab frame, in comparison.
One can see that the cell contracts and moves aside until it has ``escaped'' the activation region. 
After that it relaxes again to the stationary length without optogenetic 
perturbation, $\hat{L}$, without moving the trailing edge into the region of activation. 
Overall the cell has moved a certain fraction of its size only.

\begin{figure}[!t]
	\centering
		\includegraphics[width=\columnwidth]{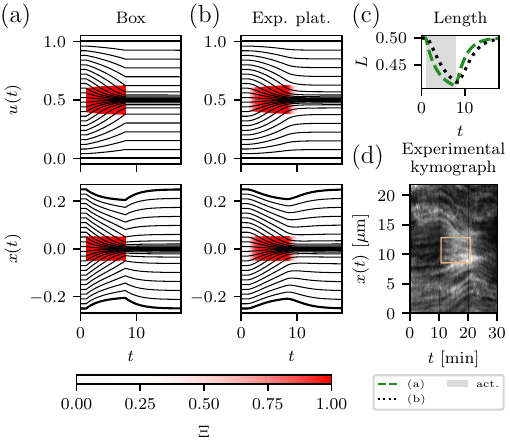}
	\caption{Kymographs for optogenetic activations with different time profiles. 
	The curves in panels (a) and (b) correspond to trajectories of material points and the 
	value of the shape function $\Xi$, corresponding to the optogenetic activation strength, is shown in color. 
	The thick lines are the cell boundaries. The upper panels in (a) and (b) show the flow in internal coordinates (u)
	and the lower ones in the lab frame (x). 
	The panels compare a temporal box signal (a) to an exponentially plateauing signal (b), 
	both with $u_\mathrm{on}(t=0)=0.4, u_\mathrm{off}(t=0)=0.6$, $t_\mathrm{on}=1, t_\mathrm{off}=8$.
	Panel (c) shows the resulting cell length and (d) an experimental kymograph along a stress fiber 
	upon optogenetic activation in the box region (modified from \cite{Oakes_2017_NatComm_Optogenetic_RhoA}).
	Parameters: $\mathcal{T}=0$, $\mathcal{L}=1$, $\sigma_{act}=0.5$, $\varepsilon=0.4$. 
	}
\label{fig:Active_Maxwell_optgen_simulation_asymm_kymograph_timeprofiles}
\end{figure}

As a last study we investigated the effects of the temporal on-off dynamics of the signal
on the internal flow, considering for simplicity again a central perturbation that can
be compared to existing experiments.
Fig.~\ref{fig:Active_Maxwell_optgen_simulation_asymm_kymograph_timeprofiles} shows
the two versions of the kymographs for a temporal box signal (a) and the, more realistic, 
exponentially plateauing activation signal (b), that was successfully used in Ref.~\cite{Oakes_2017_NatComm_Optogenetic_RhoA}.
Note that our one-dimensional model shares some general features with the model proposed there 
for the dynamics along stress fibers in fibroblasts: in Ref.~\cite{Oakes_2017_NatComm_Optogenetic_RhoA} also
an active Maxwell model was considered, but with elastic coupling to the substrate while the in-plane boundary conditions were free. 
The kymographs in Fig.~\ref{fig:Active_Maxwell_optgen_simulation_asymm_kymograph_timeprofiles}(a),(b) 
show inwards flow of the material during activation, even outside the activation region.
Panel (c) shows again the cell's length.
As soon as the signal ends, the flow is reversed and the material returns to the initial configuration. 
This is in qualitative agreement to the experimentally obtained kymographs in Ref.~\cite{Oakes_2017_NatComm_Optogenetic_RhoA}, 
as exemplified in Fig.~\ref{fig:Active_Maxwell_optgen_simulation_asymm_kymograph_timeprofiles}(d). 
Comparing the temporal profiles, we see that the activation and adaptation periods that are modeled 
with the exponential plateauing function lead to smoother changes in the trajectories, 
matching the experimental results better, but no further qualitative changes are observed.

\section{\label{sec:Polymerization} Effect of polymerization}

\subsection{\label{sec:model_polymerization}Model definition}

In motile cells one typically observes an increased actin polymerization at the leading edge, usually triggered by activation of
signaling proteins like Rac1 or Cdc42. Due to the mechanical resistance
of the cell membrane, this protrusion is partially converted into
retrograde flow. In non-motile yet spreading cells, 
both protrusive activity and retrograde flow
are symmetric, occuring along the whole cell periphery. In
order to complement our analysis by these important features,
we now consider polymerization at the boundaries
by introducing the polymerization velocities $v_p^\pm$ for the right and the left edge, respectively. We assume that the local
polymerization velocity acts in addition to the 
internal flow velocity $\partial_x\sigma(l_\pm,(t))/\xi$ from Eq.~(\ref{eq:Active_Maxwell_BC_dot_lpm}), i.e.
\begin{equation}\label{eq:Active_gel_polymerization_lpm_dot}
\dot{l}_\pm = \frac{1}{\xi}\partial_x \sigma(l_\pm(t),t) + v_p^\pm.
\end{equation}
Note that in general these kinematic boundary conditions do not conserve gel mass. 
To achieve this, one had to consider additional bulk depolymerization and conservation laws, which we neglect here,
in agreement with our assumption from above that the gel is infinitely compressible. 
We again restrict our discussion to the purely viscous case and obtain, 
cf.~Eqs.~(\ref{eq:Active_Maxwell_nondim_bvp_s_u}) and (\ref{eq:Active_Maxwell_optgen_nondim_bvp_s_u}),
the modified non-dimensionalized BVP
	\begin{equation}\label{eq:Active_gel_nondim_bvp_s_u_vp}
	\begin{gathered}
		\mathcal{L}^2 \frac{1}{L^2}\partial_{u}^2 s  - s + (L-1) = -\sigma_{act} -\varepsilon\Xi , \\
		s(u_\pm) = 0, \qquad \partial_u s(u_\pm) = L\dot{l}_\pm - Lv_p^\pm.
	\end{gathered}
	\end{equation}
	
	The equation for $s$ can be solved using the Green's function. For this we assume box-shaped temporal and spatial shape functions, where activation is between $u_\mathrm{on}$ and $u_\mathrm{off}$. Inserting this solution into the equation for $\dot{L}=\dot{l}_+-\dot{l}_-$ results in an ODE for the length, similarly to Eq.~(\ref{eq:Active_Maxwell_ode_for_length_viscous}). The equation for the general case,
	Eq.~(\ref{eq:appendix_polymerization_Ldot}),
	is given in appendix \ref{app:Polymerization}, with special cases discussed in more detail below. 
	Based on  the length in the steady state, the corresponding velocity can be determined via $V=(\dot{l}_++\dot{l}_-)/2$ and yields Eq.~(\ref{eq:appendix_polymerization_V}). 
	Importantly, the resulting length equation only depends on the polymerization velocity difference $\Delta v_p=v_p^+-v_p^-$, while the cell's center of mass velocity only depends on the average velocity $\overline{v_p}=(v_p^++v_p^-)/2$. Our results are in agreement to previous findings for a fixed length \cite{Carlsson_2011_NJP_Cell_propulsion_active_stresses, Recho_2016_MMS_Max_velocity_self_propulsion}. In the following we discuss
	instructive examples to demonstrate the effect of polymerization
	while keeping the focus on flows effected by active contraction.
	
	\subsection{\label{sec:Half_activated_polymerization_spreading}Half-activated, symmetrically spreading cell}
	First we consider a half-activated cell with $u_\mathrm{on}=0$ and $u_\mathrm{off}=1/2$. In that case we can simplify the length equation (\ref{eq:appendix_polymerization_Ldot}) to read
	\begin{equation}\label{eq:Polymerization_L_dot_halfcell}
	\dot{L}= - \frac{1}{\mathcal{L}} \Big( 2\sigma_{act} + 2(L-1) + \epsilon \Big)  \tanh(L/2\mathcal{L}) + \Delta v_p.
	\end{equation}
	The steady state equation $\dot{L}=0$ then leads to an algebraic equation determining the steady state length. Knowing this length, the velocity can be calculated according to Eq.~(\ref{eq:appendix_polymerization_V}) to be
	\begin{equation}\label{eq:Polymerization_V_halfcell}
	V=\overline{v_p} + \frac{\epsilon}{2\mathcal{L}} \tanh(L/4\mathcal{L}).
	\end{equation}
	
	To determine the steady state, we can therefore either integrate the BVP Eq.~(\ref{eq:Active_gel_nondim_bvp_s_u_vp}) using the Finite Volume Method 
	as before, or use numerical root finding methods to solve the algebraic equation for the length and insert the result into the equation for $V$. 
	
	Looking at Eqs.~(\ref{eq:Polymerization_L_dot_halfcell}) and (\ref{eq:Polymerization_V_halfcell}) one finds that the viscous length scale $\mathcal{L}$ enters as a saturation parameter for the length dependence in the $\tanh$-term. Hence it determines how far the system is in the nonlinear regime for polymerization $\Delta v_p$ and $\overline{v_p}$ added to the unperturbed steady state (without activation) $\hat{L}=1-\sigma_{act}$ and $V=0$
	and consequently determines the strength of the effect of polymerization. It does not, however, change the qualitative behavior, which was verified numerically. 
	As optogenetic activation enters in the part depending on $L$, $\mathcal{L}$ also serves as a weight determining the relative strength of polymerization vs.~optogenetic effects. This is true irrespective of the choice for $u_\mathrm{on/off}$ (c.f.~appendix~\ref{app:Polymerization}). In the following we therefore focus on the case $\mathcal{L}=1$. 
	
\begin{figure}[!t]
	\centering
	\includegraphics[width=\columnwidth]{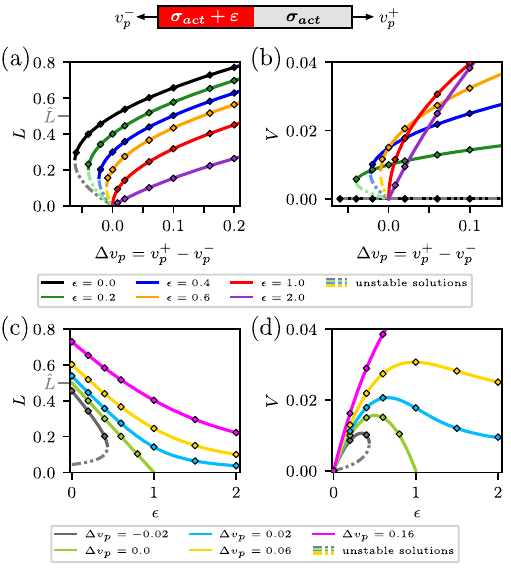}
	\caption{Shown are steady state length and velocity for the half-activated  symmetrically spreading cell (i.e.~box signal from $u_\mathrm{on}=0$ to $u_\mathrm{off}=1/2$) with opposite polymerization velocities at the trailing and leading edge, i.e. $v_p^+=-v_p^-$. 
	The sketch above depicts the one-dimensional slab with activation in red and arrows indicating the (de-)polymerization direction and strength.
	Parts (a) and (b) show the length and velocity as functions of the polymerization velocity difference $\Delta v_p$ and
	(c) and (d) as functions of the optogenetic activation strength $\varepsilon$. The symbols indicate Finite Volume simulations and the curves numerical solutions of the algebraic equations  (\ref{eq:Polymerization_L_dot_halfcell}) and (\ref{eq:Polymerization_V_halfcell})
	for $\dot{L}=0$. Stable solution branches are traced as solid, unstable ones as dash-dotted curves. 
	Other parameters:  $\mathcal{T}=0$, $\mathcal{L}=1$, $\sigma_{act}=0.5$.}
\label{fig:Active_Maxwell_polymerization_half_actived_spreading}
\end{figure}	

For a symmetrically spreading cell, the polymerization velocities point outwards at both edges with equal magnitude, i.e. $v_p^+=-v_p^-$, corresponding to the non-polarized non-motile state of a cell with retrograde actin flow. While $\overline{v_p}=0$, we have a positive polymerization velocity difference $\Delta v_p$. From Eq.~(\ref{eq:Polymerization_L_dot_halfcell}) one expects that the length increases for $\Delta v_p>0$. Fig.~\ref{fig:Active_Maxwell_polymerization_half_actived_spreading} shows the steady state length $L$ and velocity $V$ as functions of $\Delta v_p$ and the activation strength $\varepsilon$. We see that indeed the length increases with larger polymerization velocity differences. For negative $\Delta v_p$ we have the opposite effect, until the spring and anterograde actin flow are not able to balance both the contractile active stress and the depolymerization and the cell collapses, resulting in a saddle-node bifurcation, c.f.~Fig.~\ref{fig:Active_Maxwell_polymerization_half_actived_spreading}(a). However, this case is biologically not really relevant, as such anterograde flow has only been observed at concave cell edges \cite{Chen_2019_Nature_Curvature_sensing_directional_actin_flow_migration}, which are not found at both edges along the motility axis in single-cell motility experiments. 

One also sees that increasing the optogenetic activation strength $\epsilon$ leads to additional contractions, as observed earlier. 
The velocity of the cell, which is positive due to the activation of the left part of the cell, increases as $\Delta v_p$ increases, indicating that internal flow can strengthen contractile motility initiation, even if no symmetry breaking in the polymerization is assumed, c.f.~Fig.~\ref{fig:Active_Maxwell_polymerization_half_actived_spreading}(b). This can be attributed to the increased length, which implies a larger integrated optogenetic stress in lab coordinates.

The dependence of the steady state length on the optogenetic activation strength $\varepsilon$, shown in Fig.~\ref{fig:Active_Maxwell_polymerization_half_actived_spreading}(c), changes qualitatively as $\Delta v_p$ switches its sign. For edge depolymerization we see cell collapse for finite $\varepsilon$. For symmetrical polymerization we find that the length does not collapse anymore, no matter how large the additional asymmetrical contractile stress is. This means that even for large perturbations of the contractility the cell length remains stable. This holds true also if the additional asymmetric contractile stress is not interpreted as an optogenetic signal but rather as a two-compartment model for cell motility, where the activation region has an increased myosin concentration. 

Overall this suggests the conclusion that symmetrical polymerization at both edges could be employed to assure cell stability and is able to do so even for large contraction driven asymmetries. This represents an alternative stabilization mechanism to the previously proposed dominance of the effective elastic constraint \cite{Putelat_2018_PRE_stress_regulator_cell_motility}. Note that this also could
settle the issue of cell collapse occurring frequently in similar models with asymmetric spatially-dependent myosin concentration fields, as an alternative to nonlinear elastic coupling \cite{recho_contraction-driven_2013, Putelat_2018_PRE_stress_regulator_cell_motility, Recho_2015_JMPS_Mechanics_motility_initiation_arrest}. 

Making a connection to section \ref{sec:Active_Maxwell_optgen},
the there-discussed case of $\Delta v_p=0$ is special and marks the transition from instability to stability for increased activation strengths. Based on Eq.~(\ref{eq:Polymerization_V_halfcell}) we see that the stable length is given by $L=1-\sigma_{act}-\varepsilon/2$, which verifies the previously found approximation $\hat{L}_\mathrm{ap}$,  Eq.~(\ref{eq:Active_Maxwell_optgen_approx_equation_hat_L}), for the special case of $\Xi$ leading to a vanishing integral over the stress field without the boundary conditions $s$.

Coming back to Fig.~\ref{fig:Active_Maxwell_polymerization_half_actived_spreading},
we find that the velocity is non-monotonous as $\varepsilon$ increases, cf.~Fig.~\ref{fig:Active_Maxwell_polymerization_half_actived_spreading}(d). 
This is a consequence of the competition of the increasing activation strength $\varepsilon$ and the concomitantly decreasing length $L$ in (\ref{eq:Polymerization_V_halfcell}) and can be explained by using the integrated optogenetic contractile stress in lab coordinates, i.e.
\begin{equation}
\varepsilon L \int_{u_\mathrm{on}}^{u_\mathrm{off}}\Xi(u)\mathrm{d}u = \frac{\varepsilon L }{2} = \frac{\varepsilon (1-\sigma_{act}-\epsilon/2)}{2},
\end{equation}
where in the last step, we considered the steady state length for $\Delta v_p=0$.
The length decreases as $\varepsilon$ increases, which leads to a decreasing integrated contractile stress. In the case of $\Delta v_p=0$ this yields a downwards parabola, 
with zeros at $\varepsilon=0$ and $\varepsilon=2-2\sigma_{act}$, similar to the velocity $V$ shown in Fig.~\ref{fig:Active_Maxwell_polymerization_half_actived_spreading}(d).

\begin{figure}[!t]
	\centering
	\includegraphics[width=\columnwidth]{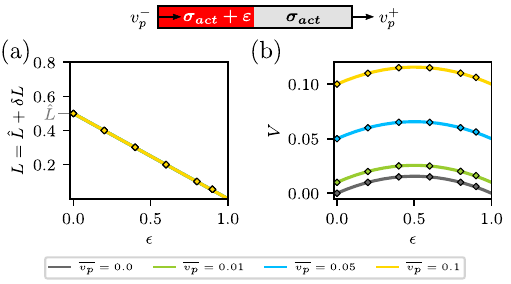}
	\caption{Shown are steady state length and velocity for the case of a half-activated 
	symmetrically moving cell with equal de-/polymerization velocities at the trailing and leading edge, respectively, 
	i.e.~$v_p^+=v_p^-$. The sketch depicts the one-dimensional slab with activation in red and arrows indicating the (de-)polymerization direction and strength.
	Parts (a) and (b) depict the length and velocity as functions of the optogenetic activation strength $\varepsilon$. The symbols indicate Finite Volume simulations and the curves numerical solutions of the algebraic equations  (\ref{eq:Polymerization_L_dot_halfcell}) and (\ref{eq:Polymerization_V_halfcell})
	for $\dot{L}=0$. 
	Other parameters:  $\mathcal{T}=0$, $\mathcal{L}=1$, $\sigma_{act}=0.5$.}
\label{fig:Active_Maxwell_polymerization_half_actived_moving}
\end{figure}	

\subsection{\label{sec:Half_activated_polymerization_moving}Half-activated, symmetrically moving cell}

Next we consider the case of a symmetrically moving cell with equal depolymerization and polymerization velocities at the trailing and leading edge, $v_p^+=v_p^-$, as proposed 
previously \cite{Kruse_2005_EPJE_Theory_Active_Polar_Gels_Paradigm_Cytoskeleton,
Juelicher_2007_PhyRep_Review_Active_Behavior_Cytoskeleton,
Kruse_2018_LectureNotes_Hydrodynamics_Active_Cytoskeleton}. 
This implies $\Delta v_p=0$ and $\overline{v_p}=v_p^+=v_p^-$. According to Eqs.~(\ref{eq:Polymerization_L_dot_halfcell}) and (\ref{eq:Polymerization_V_halfcell}) we 
therefore expect that the length now only depends on $\varepsilon$, as in the previous case for $\Delta v_p=0$, and not on $\overline{v_p}$, while the velocity is offset 
by $\overline{v_p}$. 
This was verified numerically. 
Fig.~\ref{fig:Active_Maxwell_polymerization_half_actived_moving}(a) shows that the steady state length is identical to the one found in the symmetrically spreading 
case for $\Delta v_p=0$ (cf.~the green curve in 
Fig.~\ref{fig:Active_Maxwell_polymerization_half_actived_spreading}(c). Fig.~\ref{fig:Active_Maxwell_polymerization_half_actived_moving}(b) shows  
that the velocity still has a parabolic
dependence in $\varepsilon$
with an offset given by $\overline{v_p}$.

Overall we find that the influence of polymerization is decoupled into the antisymmetric component $\Delta v_p$ influencing the length $L$ 
and the symmetric component $\overline{v_p}$
adding an offset to the velocity $V$. 
Importantly, this is the case for 
arbitrary signal shapes and positions. 

Let us now discuss the interesting question whether an optogenetic signal can be used to stop a moving cell. 
Using an asymmetric activation at the leading edge of a moving cell,
the movement can in fact be arrested for 
a signal tailored to the polymerization velocity if $\overline{v_p}$ is not too large.

We can make this statement more precise by considering the example of a half-activated symmetrically moving cell 
with $\overline{v_p}<0$ (i.e.~moving to the left)
and $\Delta v_p=0$
with activation region from $u_\mathrm{on}=0$ to $u_\mathrm{off}=1/2$ (i.e.~the cell half close to the leading edge). 
We first determine 
$\varepsilon_\mathrm{max}$, which 
is the activation strength that yields the maximum velocity $V$.
In Fig.~\ref{fig:Active_Maxwell_polymerization_half_actived_moving}(b) this corresponds to   $\varepsilon_\mathrm{max}\approx 1/2$;
in general it is given by the solution of 
\begin{equation}\label{eq:Active_gel_polymerization_moving_epsmax}
\sinh\left(\frac{1-\sigma_{act}-\varepsilon_\mathrm{max}/2}{2\mathcal{L}}\right)  = \frac{\varepsilon_\mathrm{max}}{4\mathcal{L}}.
\end{equation}
Then motility can be stopped
if $|v_p|$ is smaller than the optogenetic part of $V$ (i.e.~the one proportional to $\varepsilon$ 
in Eq.~(\ref{eq:Polymerization_V_halfcell}))
at this maxmimum  $\varepsilon_\mathrm{max}$, or
\begin{equation}\label{eq:Polymerization_vpbar_limit}
-\frac{\varepsilon_\mathrm{max}}{2\mathcal{L}} < -\frac{\varepsilon_\mathrm{max}}{2\mathcal{L}}\tanh\left(\frac{1-\sigma_{act}-\varepsilon_\mathrm{max}/2}{4\mathcal{L}}\right) \leq \overline{v_p} \leq 0.
\end{equation}
The arrest of cell movement therefore depends 
on the viscous length scale $\mathcal{L}=\sqrt{\eta/(\xi L_0^2)}$, with larger $\mathcal{L}$ reducing the maximum stoppable polymerization velocity $\overline{v_p}$. This agrees with the previous discussion of this length scale being a weighting factor determining the relative strength of optogenetic vs.~polymerization effects.

To assess  whether an arrest of a motile cell
is within experimentally accessible ranges,
let us briefly estimate the relevant parameters.
The viscosity and active stress can be estimated from \cite{Recho_2015_JMPS_Mechanics_motility_initiation_arrest, Barnhart_2011_PLOS_Adhesive_dependent_mechanisms_motile_cell_shape, Oakes_2017_NatComm_Optogenetic_RhoA} to be $\eta=10^5$\,Pa\,s and $\sigma_{act}=10^3$\,Pa. 
The typical size of a keratocyte is $L_0=20\,\mu$m. 
The drag coefficient can vary depending on the substrate, for a medium adhesion strength
\cite{Barnhart_2011_PLOS_Adhesive_dependent_mechanisms_motile_cell_shape} we estimate 
$\xi = 2\cdot10^{14}$\,Pa\,s\,$/{\rm m}^{2}$.  
The cortex stiffness can be estimated to be $k=10^4$\,Pa \cite{Recho_2015_JMPS_Mechanics_motility_initiation_arrest, Loosley_2012, Barnhart_2010}. These parameters imply 
$\mathcal{L}^2\simeq1.25$ and $\sigma_{act}\simeq0.1$. The maximum activation can then be determined from (\ref{eq:Active_gel_polymerization_moving_epsmax}) to be $\varepsilon_\mathrm{max}\approx0.9$, which is slightly larger than the experimentally found value, which is 
between $\varepsilon=0.1$ to $0.5$ \cite{Oakes_2017_NatComm_Optogenetic_RhoA}. Taking $\varepsilon=0.5$, we arrive for the estimate of the maximal stoppable average polymerization velocity at
$|\overline{v_p}| \lessapprox 81\ \mathrm{nm}\ \mathrm{s}^{-1}$,
which is only slightly smaller than the experimentally found polymerization velocity of actin (of the order of hundreds of nanometers per second \cite{Mogilner_1996, Mogilner_2003}). Note that this maximal velocity could be larger in experimental realizations, due to the increasing effect of outwards polymerization, i.e. $\Delta v_p>0$, on the optogenetically induced velocity, cf.~Fig.~\ref{fig:Active_Maxwell_polymerization_half_actived_spreading}(d).

\subsection{\label{sec:Half_activated_polymerization_asymm}Half-activated, asymmetrically moving cells}

\begin{figure}[!t]
	\centering
		\includegraphics[width=\columnwidth]{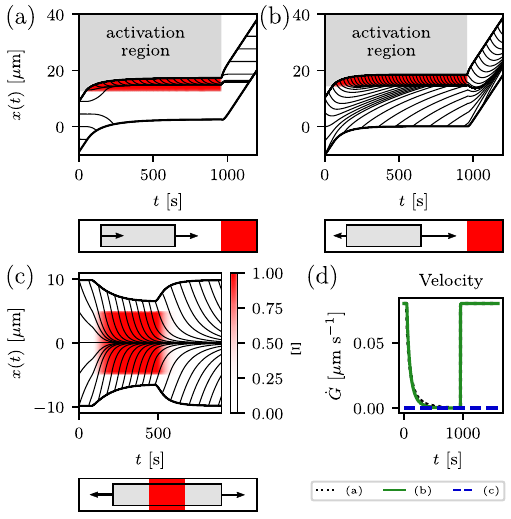}
	\caption{Kymographs for the model with optogenetic activation and polymerization. The three different model
	variants differ in the polymerization velocities (black arrows) and location of the activation region (red)
	as shown schematically below the plots of the material points (cell boundaries as thick black lines). 
	In (a) the symmetrically and in (b) the asymmetrically moving cases with a stopping optical signal region are shown, with $v_p^\pm=80$~nm~s$^{-1}$ and $v_p^+=200$~nm~s$^{-1}$, $v_p^-=-40$~nm~s$^{-1}$, respectively. In (c) the symmetrically spreading case with a centered signal with exponentially plateauing temporal shape function is shown with $v_p^\pm=\pm80$~nm~s$^{-1}$. 
	In (d) the velocites are shown for the three cases. The parameters correspond to the experimentally relevant parameters given in the text, i.e.~$\mathcal{L}^2=1.25$, $\sigma_{act}=0.1$, $\varepsilon=0.5$, $\mathcal{T}=0$.
	}
	\label{fig:Active_Maxwell_polymerization_kymographs}
\end{figure}

Asymmetric polymerization velocities  occur e.g.~during the transition into a polarized moving state 
or due to additional regulatory processes.
For outwards pointing $v_p^\pm$ we expect to find a superposition of the length effect, caused by $\Delta v_p$, and the velocity effect, caused by $\overline{v_p}$. 
Indeed, we found that $\overline{v_p}$ leads to an offset of the velocities (c.f.~Fig.~\ref{fig:Active_Maxwell_polymerization_half_actived_moving}), while the behavior with respect to $\Delta v_p$ is unchanged (c.f.~Fig.~\ref{fig:Active_Maxwell_polymerization_half_actived_spreading}). 
Numerically we find an overdamped relaxation towards the motile steady state. The asymmetrical polymerization velocities are compensated for by asymmetrical flow patterns of the actomyosin network, as visible in 
Fig.~\ref{fig:Active_Maxwell_polymerization_kymographs}(b), leading to a stable length and velocity. This shows the consistency of our results, irrespective of the assumptions made on the boundary polymerization.

To check the arrest of motion via optogenetics predicted in the previous section
and whether it is experimentally accessible, 
we performed Finite Volume simulations for the previously described experimentally relevant parameters for symmetrically moving and asymmetrically spreading cells.
Results are shown in
Fig.~\ref{fig:Active_Maxwell_polymerization_kymographs}(a) and (b), where we introduced an activation region which lies outside the motile cell and blocks the way in the direction of polymerization-driven movement.
For the realistic parameters we found that in both cases the locomotion is arrested throughout the period of activation and recommences after turn-off. 

Upon increasing the average polymerization velocity $\overline{v_p}$, first the symmetrically moving cell cannot be stopped anymore and then even asymmetrically moving ones. The velocity decreases until the edge of the activation region aligns with the cell center, corresponding to the analytically studied situation, and then increases again. The fast velocity relaxation after turn off, as evident from Fig.~\ref{fig:Active_Maxwell_polymerization_kymographs}(d), indicates that the time dependence is dominated by the movement of the activation region's edge and not by the cell's adaptation, which implies that the estimated limit Eq.~(\ref{eq:Polymerization_vpbar_limit}) is attained as upper boundary for the stoppable $|\overline{v_p}|$ for $\Delta v_p=0$. 

\subsection{\label{sec:Center_activated_polymerization_spreading}Center-activated, symmetrically spreading cell}

Finally we considered a centered activation with the same integrated signal (in internal coordinates) as in the previous case, 
i.e.~$u_\mathrm{on}=1/4$ and $u_\mathrm{off}=3/4$. From the previous discussion we know that the length and velocity decouple from the polymerization
and that the length is primarily given by subtracting the integral of the optogenetic activation stress from the unperturbed stable length $\hat{L}$. The equations for length  and velocity, Eqs.~(\ref{eq:appendix_polymerization_Ldot}) and (\ref{eq:appendix_polymerization_V}), now read
\begin{multline}\label{eq:Active_gel_polymerization_centered_Ldot}
\dot{L}= - \frac{1}{\mathcal{L}} \Big( 2\sigma_{act} + 2(L-1)  \Big)  \tanh(L/2\mathcal{L}) + \Delta v_p \\-\frac{2\varepsilon}{\mathcal{L}} \frac{\cosh\left(3L/4\mathcal{L} \right) - \cosh\left(L/4\mathcal{L} \right)}{\sinh\left(L/\mathcal{L}\right)},
\end{multline}
and we simply have $V=\overline{v_p}$.
The last term in  Eq.~(\ref{eq:Active_gel_polymerization_centered_Ldot}), which accounts for the optogenetic activation, shows only a relative deviation to the corresponding term in the asymmetrically half-activated cell of 3\% for realistic parameters and $|L/\mathcal{L}| \lessapprox 1$.
We therefore obtain similar results for the lengths as in the half-activated cell. However, we do not have any symmetry breaking in the activation, as the velocity is solely determined by the average polymerization velocity $\overline{v_p}$ and does no longer depend on $\Delta v_p$ and $\varepsilon$.
The respective kymograph for a signal fixed in co-moving coordinates is shown in 
Fig.~\ref{fig:Active_Maxwell_polymerization_kymographs}(c).

\section{\label{sec:Active_Maxwell_summary}Summary and conclusions}

Here we have analyzed a mathematical framework that allows us to
study the effect of optogenetic activation of contractility on cell spreading and migration. The cell is modeled as an active gel
with Maxwell viscoelasticity, reflecting the viscous nature
of the actomyosin cytoskeleton on long time scales. However, by
using elastic boundary conditions, we also represent the fact
that cells have a typical size that is controlled by 
other processes not explicitly included here (e.g. volume
control by ion channels). Optogenetic activation was introduced 
in the governing equation for the stress as a localized additional
contribution to the active stress. While the spatial part of the 
activation profile corresponds to the typical laser profiles
used in experiments, the temporal part is set by the
reaction-diffusion system of the used optogenetic construct. 
Here we have used generic shapes which in the future could
be replaced by experimentally measured transfer functions.

In the basic version without optogenetic activation and
without polymerization, the proposed simple model has only one steady state, which is the resting state, i.e.~non-motile.
Upon perturbation, the contractile model has different time scales for length relaxation (slow) 
and for re-symmetrization of internal stress/flow profiles (fast). 
In the purely viscous limit,  re-symmetrization is instantaneous  (as the analytical solution is symmetric). 
In the viscoelastic case, the re-symmetrization timescale is determined by the ratio of the renormalized Maxwell relaxation time and the squared relative length scale,  
$\mathcal{T}/\mathcal{L}^2$, i.e.~the only two dimensionless parameters. 
The model describes generic features of spreading cells well, but is too simple to
describe spontaneous motility. 

Optogenetic perturbations can be applied both in lab
and co-moving frames and we discussed both cases throughout this work.
Since no motile steady state exists in our basic model, 
after optogenetic activation has ceased the cell relaxes back to the only existing steady state, which is the resting one. 
Our study of various shape functions gave the following general results:
The specific spatial shape function does not matter for the cell's length adaptation,
only the integrated signal is important.
The asymmetry of the signal does not matter for the length change, but it does matter 
for the transient motility: the closer to the boundary the perturbation, the faster the cell moves. 
For the temporal shape function, the exponential protocol yields best results
when compared to experiments. 
It is possible to use simpler shape functions, but care has to be taken 
for a consistent modeling in case
the adaptation time to the signal exceeds the experimental activation time.

The only way to achieve persistent motility in the contraction model
is to use asymmetric optogenetic activation
in the co-moving frame, which should be experimentally realizable. 
When using an asymmetric activation
that is stationary in the lab frame, the cell escapes the activation signal and then settles down.
Upon symmetric (central) activation, the internal flow profiles are dominated by the length change with additional flow towards the activation region. This finding agrees qualitatively with experimental results for dynamics of stress fibers in fibroblasts \cite{Oakes_2017_NatComm_Optogenetic_RhoA}.

In order to increase complexity and include an important biological
effect that usually is also present in cells, we also 
considered the effect of polymerization. Because here we
focus on the role of contractility, we included 
polymerization as a boundary effect; future work 
could also study bulk effects of polymerization, but
our approach seems appropriate since actin polymerization for cell migration is
usually effected by signaling processes at the plasma membrane.
In our theory, the effects of polymerization can be decomposed into the polymerization velocity difference (the outwards vs.~inwards pointing antisymmetric part of the velocities) and the average polymerization velocity (velocities pointing in the same directions). The velocity difference couples to the length and outwards polymerization stabilizes the solution such that large optogenetic activations still yield positive lengths. Since the activation can also be interpreted as an inhomogeneous contractile stress resulting from cell-internal processes, this suggests a general stabilization of cell shape through polymerization. 
The average velocity, in turn, leads to an offset adding to the total cell velocity. 

For a symmetrically spreading cell we find that optogenetic activation can induce motility. It also can control polymerization-driven motility and it can arrest moving cells in a parameter range that can be estimated analytically and that includes the experimentally relevant parameters. These results suggest that it is an attractive 
experimental strategy to affect cell migration in a situation in which the cell already shows protrusive activity. 

To conclude, the simple model discussed here allows for detailed studies of the effect of optogenetic perturbations on spreading and moving cells.
However, in the absence of polymerization, cells in the model cannot maintain a broken symmetry on their own and there is no bistability of resting and moving states. Therefore, it would be interesting to study -- along the same lines as developed here --
more complex models that include internal degrees of freedom, e.g.\ the spatial concentration profile of motors
\cite{Recho_2015_JMPS_Mechanics_motility_initiation_arrest} or nonlinearities, because both would allow the cell to maintain a broken symmetry.
A motor concentration field would also allow us to separate the effects of having more 
motors due to assembly and having higher levels of motor activity, a distinction
that cannot be made in our simple model. Interestingly, however, this distinction 
can also not necessarily be made in optogenetic experiments with the Rho-pathway, which
activates both myosin II minifilament assembly and myosin II motor head cycling.

Another interesting direction for future work is the consideration of dimensionality. Optogenetics explicitly allows for controlled spatial activation and therefore flows could also be controlled
in two or even three dimensions. For instance, experimentally a curvature dependence of the actin flow at the cell edge has been observed \cite{Chen_2019_Nature_Curvature_sensing_directional_actin_flow_migration}.
Extensions of active gel models with myosin concentrations \cite{Berlyand_2017_arxiv_bifurcation_traveling_waves_Keller_Segel,Berlyand_2019_Stability_steady_states_traveling_waves_free_boundary} to two dimensions have already been proposed and now could be used
to explore the effect of non-trivial spatial profiles. In general,
our approach can be used also for the inverse problem of 
predicting appropriate illumination patterns for a desired
kind of flow or movement, and this not only for cells, but
also for biomimetic systems like synthetic cells.

\begin{acknowledgments}

This research was conducted within the Max Planck School Matter to Life supported by the German Federal Ministry of Education and Research (BMBF) in collaboration with the Max Planck Society. We also acknowledge support  
by the Deutsche Forschungsgemeinschaft (DFG, German Research Foundation) – Projektnummer 390978043.
USS is member of the Interdisciplinary Center for Scientific Computing (IWR) at Heidelberg.

\end{acknowledgments}

%
%
%
\appendix

\section{\label{appendix_lack_motile_steadstate} Lack of motile steady states}




Introducing $y_1(u) = (L/\mathcal{L})s(u)$ and $y_2(u)=\partial_u s(u) - VL$, Eq.~(\ref{eq:Active_Maxwell_nondim_bvp_s_u}) can be rewritten in two-dimensional phase space to read
\begin{equation}\label{eq:Lack_motile_steady_state_dyn_sys}
\begin{aligned}
\partial_u y_2 =& + \frac{\mathcal{T}}{\mathcal{L}^2} (VL) y_2 + \frac{\mathcal{T}}{\mathcal{L}^2} (y_2)^2 +  \frac{L}{\mathcal{L}} y_1 \\ &\quad- \frac{L^2}{\mathcal{L}^2}  (L-1) -  \frac{L^2}{\mathcal{L}^2} \sigma_{act}, \\
\partial_u y_1 =& \frac{L}{\mathcal{L}} y_2 + \frac{L}{\mathcal{L}} VL.
\end{aligned}
\end{equation}
The boundary conditions then imply $y_1(0)=y_1(1)=0$ and $y_2(0)=y_2(1)=0$. For $V\neq0$ we only have one fixed point, which is not at $(0,0)$. This means that a solution to the original problem must correspond to a periodic orbit with period 1, starting and ending at  $(0,0)$.

However, Eq.~(\ref{eq:Lack_motile_steady_state_dyn_sys}) is a gradient system with potential
\begin{multline}\label{eq:Lack_motile_steady_state_potential}
V(y_1, y_2)=-\frac{\mathcal{T}}{\mathcal{L}^2} (VL) \frac{1}{2}(y_2)^2 - \frac{\mathcal{T}}{\mathcal{L}^2} \frac{1}{3} (y_2)^3 - \frac{L}{\mathcal{L}} y_1 y_2 \\- \frac{L}{\mathcal{L}}  (VL) y_1 + \frac{L^2}{\mathcal{L}^2}  \Big[(L-1)+\sigma_{act}\Big] y_2,
\end{multline}
and hence cannot have any closed orbits \cite{Strogatz_2015_Book_Nonlinear_dynamics}, thus proving the nonexistence of motile steady states.  Note that $V(y_1,y_2)$ without the cubic $(y_2)^3$ term constitutes a potential for the linear Maxwell model.

\section{\label{appendix_numerics} Numerical implementation} 

We introduce the rescaled stress field $\hat{\sigma}=L\sigma$ into the BVP, Eq.~(\ref{eq:Active_Maxwell_optgen_nondim_bvp})
, and transform into internal coordinates. 
By denoting the cell's center by $G=(l_++l_-)/2$ and defining the advection velocity field 
\begin{equation}
\hat{v}(u,t)=\frac{\dot{G}(t)}{L(t)}-\frac{\dot{L}(t)}{L(t)}\left(\frac{1}{2}-u\right),
\end{equation}
Eq.~(\ref{eq:Active_Maxwell_optgen_nondim_bvp}) 
becomes an advection-diffusion equation with source term
for the stress field $\hat{\sigma}$: 
\begin{multline}\label{eq:Active_Maxwell_numerical_convection_advection}
\mathcal{T}\partial_t\hat{\sigma} = \partial_u \left[ \left(\frac{\mathcal{L}^2}{L^2} + \frac{\mathcal{T}}{L^2} \varepsilon \Xi\right) \partial_u \hat{\sigma} \right] \\+ \partial_u \left[ \left(\hat{v} - \frac{\mathcal{T}}{L^3} \partial_u\hat{\sigma} \right) \hat{\sigma} \right]
 - \mathcal{T} L (\hat{v} \varepsilon \partial_u \Xi) + \mathcal{T} L (\varepsilon \partial_t \Xi) \\+ \frac{\mathcal{T}}{L^3} \left[ \hat{\sigma} \partial_u^2 \hat{\sigma} - \varepsilon \Xi L \partial_u^2 \hat{\sigma} \right]  - \hat{\sigma} + \sigma_{act} L + \varepsilon L \Xi.
\end{multline}
The boundary condition for $\hat{\sigma}$ and the time evolution of $L$ and $G$
are given by
\begin{subequations}	
\begin{align}\label{eq:Active_Maxwell_numerical_convection_advection_bqs_a}
	&\hat{\sigma}(0,t) = \hat{\sigma}(1,t) = - L(L-1), \\
	\label{eq:Active_Maxwell_numerical_convection_advection_bqs_b}
	&\dot{L}(t) = \dot{l}_+(t) - \dot{l}_-(t) = \frac{\partial_u \hat{\sigma}(1,t) - \partial_u \hat{\sigma}(0,t)}{L^2} , \\
	\label{eq:Active_Maxwell_numerical_convection_advection_bqs_c}
	&\dot{G}(t) = \frac{\dot{l}_+(t) + \dot{l}_-(t)}{2} = \frac{\partial_u \hat{\sigma}(1,t) + \partial_u \hat{\sigma}(0,t)}{2L^2}.
\end{align}
\end{subequations}

Eq.~(\ref{eq:Active_Maxwell_numerical_convection_advection}) was solved 
using the Finite Volume Method, using FiPy \citep{Guyer_2009_CompSciEng_FiPy}. 
$\hat{\sigma}$ was discretized on a regular mesh with $n_\mathrm{mesh}=50$ mesh points 
without optogenetic signals and an increased spatial resolution of $n_\mathrm{mesh}=400$ 
for simulations with optogenetic perturbations. The nonlinearities in Eq.~(\ref{eq:Active_Maxwell_numerical_convection_advection}) were solved iteratively by solving the linear Finite Volume system with inserted nonlinear coefficients until the residual vector of our solution from the nonlinear part had a norm $<10^{-8}$ \cite{Guyer_2009_CompSciEng_FiPy}.

Given the initial data,
equations (\ref{eq:Active_Maxwell_numerical_convection_advection}) 
and (\ref{eq:Active_Maxwell_numerical_convection_advection_bqs_a}) 
were integrated using the upwind Finite Volume scheme \citep{Balachandar_2018_Book_Computational_Fluid_Dynamics}. 
Based on the resulting stress field, (\ref{eq:Active_Maxwell_numerical_convection_advection_bqs_b}) 
and (\ref{eq:Active_Maxwell_numerical_convection_advection_bqs_c}) were then integrated using  Euler stepping.

\section{\label{app:Shape_Functions}Formulas of different shape functions}

In the following we assume a spatial extension of the signal given by $ u_{\textrm{on}},u_{\textrm{off}} \in [0,1]$ 
in internal coordinates at time $t=0$ for all formulas.

A shape function in the cell's co-moving system moves with the center of the cell, $G$, 
but stays constant in width in the lab frame. 
Defining 
\begin{equation}
a_{{\textrm{on}}/{\textrm{off}}}=\frac{L(0)}{L(t)}\left( u_{\textrm{on}/\textrm{off}} - \frac{1}{2}\right) + \frac{1}{2}, 
\end{equation}
and using the characteristic function $\chi_A$ for set $A$ \footnote{
The characteristic function $\chi_A$  is given by $\chi_A(x)=1$ for $x\in A$ and $\chi_A(x)=0$ otherwise.},
a co-moving box signal then reads
\begin{equation}\label{eq:appendix_shape_functions_Box_cm}
  \Xi^{\textrm{cm}}_{\textrm{box}}(u,t) 
  = \chi_{ \left[ a_{\textrm{on}},a_{\textrm{off}}  \right] }(u),
  \end{equation}
in internal coordinates $u$. Note that $a_{{\textrm{on}}/{\textrm{off}}}$ arises when
the original $u_{\textrm{on}/\textrm{off}}\in[0,1]$ is transformed into 
the co-moving frame by $x_{\textrm{on}/\textrm{off}}(t=0)-G(0)=L(0)(u_{\textrm{on}/\textrm{off}}-1/2)$, 
then moved along with the cell as $x_{\textrm{on}/\textrm{off}}(t)=x_{\textrm{on}/\textrm{off}}(0)-G(0)+G(t)$ 
and finally mapped back into the $u$-frame.
  
A signal that stays constant in the lab system (i.e.~in the $x$ coordinates) 
moves in internal coordinates a (rescaled) distance
\begin{equation}
g(t)=\frac{G(t)-G(0)}{L(t)} 
\end{equation}
during time $t$ and hence the signal
reads
  \begin{equation}\label{eq:appendix_shape_functions_Box_x}
  \Xi^{\textrm{x}}_{\textrm{box}}(u,t) = 
  \chi_{\left[  a_{\textrm{on}} - g(t),  a_{\textrm{off}} - g(t) \right]}(u). 
  \end{equation}
 
Since experimental laser spots are not box functions, we also studied smooth
Gaussian signals with center $\mu\in[0,1]$ and width $\omega$ in internal coordinates. 
The normalizations are chosen such that the integral $\int\Xi_s(u,t)\mathrm{d}u$ of the Gaussian signal 
is identical to the one of a box function with $\omega=(u_\textrm{off}-u_\textrm{on})/2$. 

Defining the rescaled center and width,
  \begin{equation}
  \tilde{\mu}=\frac{L(0)}{L(t)}\left(\mu-\frac{1}{2}\right)+\frac{1}{2},\quad \tilde{\omega}=\frac{L(0)}{L(t)}\omega,
  \end{equation}
   respectively, the resulting shape functions are
     \begin{equation}
  \begin{aligned}\label{eq:appendix_shape_functions_Gaussian_x}
   \Xi^{\textrm{cm}}_{\textrm{Gauss}}(u,t)&\equiv\Xi^{\textrm{cm}}_{\textrm{Gauss}}(u,t;\ \tilde{\mu}) = \sqrt{\frac{2}{\pi}} \exp\left(-\frac{\left(u-\tilde{\mu}\right)^2}{2\tilde{\omega}^2}\right),\\
  \Xi^{\textrm{x}}_{\textrm{Gauss}}(u,t) & 
  = \Xi^{\textrm{cm}}_{\textrm{Gauss}}\left(u,t;\ \tilde{\mu}-g(t)\right).
  \end{aligned}
  \end{equation}   
   
For the temporal shape functions, with turn on/turn off times $t_{\mathrm{on}}$ and $t_{\mathrm{off}}$, respectively, 
the box function reads
  \begin{equation}\label{eq:appendix_shape_functions_t_box}
  \Xi_t^{\textrm{box}}(t)= \chi_{[t_{\mathrm{on}},\ t_{\mathrm{off}}]}(t).
  \end{equation}
We also studied a continuous version of the box signal, using smooth $\tanh$-functions
  \begin{multline}\label{eq:appendix_shape_functions_t_tanh}
  \Xi_t^{\mathrm{tanh}}(t) = \frac{1}{2} \tanh\left(\frac{2 (t-t_{\mathrm{on}})}{\alpha (t_{\mathrm{off}}-t_{\mathrm{on}})}\right) \times \\ \tanh\left(\frac{2 (t_{\mathrm{off}}-t)}{\alpha (t_{\mathrm{off}}-t_{\mathrm{on}})}\right) + \frac{1}{2},
  \end{multline}
where $\alpha$ ($\alpha=0.1$ for the given parameters) determines the sharpness and was chosen such  
that the shape function is almost at 1 respectively 0 at 
$t=t_\mathrm{on} \pm \alpha (t_\mathrm{on}-t_{\mathrm{off}})$ 
and $t=t_\mathrm{off} \mp \alpha (t_\mathrm{on}-t_{\mathrm{off}})$. 
Finally, as suggested by Ref.~\cite{Oakes_2017_NatComm_Optogenetic_RhoA} we also considered
  \begin{multline}\label{eq:appendix_shape_functions_t_exp}
  \Xi_t^{\mathrm{exp}}(t) = \left(1-\exp\left(-\frac{t-t_\mathrm{on}}{\alpha(t_\mathrm{off}-t_\mathrm{on})}\right)\right) \chi_{[t_\mathrm{on},t_\mathrm{off})}(t) +\\ \left(1-\exp\left(-1/\alpha\right)\right) \exp\left(-\frac{t-t_\mathrm{off}}{\alpha(t_\mathrm{off}-t_\mathrm{on})}\right) \chi_{[t_\mathrm{off},\infty)}(t),
  \end{multline}
  which is an exponential plateauing function with (de-) activation time $\alpha(t_\mathrm{off}-t_\mathrm{on})$, 
  implementing 
  that activation and deactivation have approximately the same time constants (here we take $\alpha=0.2$). 

\section{\label{app:Polymerization}Full equations for length and velocity  with activation and polymerization}
Based on the solution of (\ref{eq:Active_gel_nondim_bvp_s_u_vp}) we can derive the ODE for the length relaxation for a shape function of $\Xi=\chi_{[u_\mathrm{on},u_\mathrm{off}]}$. The activation results in an additional term from integrating the Green's function in the activation region and evaluating the result at the boundary. We introduce
\begin{equation}
I(\zeta_0, \zeta_1) \equiv \frac{\cosh\left(\frac{L}{\mathcal{L}} \zeta_0\right) - \cosh\left(\frac{L}{\mathcal{L}} \zeta_1\right)}{\sinh(\frac{L}{\mathcal{L}})}.
\end{equation} 
Similarly to (\ref{eq:Active_Maxwell_ode_for_length_viscous}), we obtain for the length
\begin{multline}\label{eq:appendix_polymerization_Ldot}
\dot{L}= - \frac{1}{\mathcal{L}} \Big( 2\sigma_{act} + 2(L-1) \Big)  \tanh(L/2\mathcal{L}) + \Delta v_p \\+ \frac{\epsilon}{\mathcal{L}}\ \Big( I(u_\mathrm{on},\ u_\mathrm{off}) -   I\big(1-u_\mathrm{on},\ 1-u_\mathrm{off}\big) \Big),
\end{multline}
where we introduced the polymerization velocity difference $\Delta v_p= v_p^+-v_p^-$. If we know the steady state length, we can calculate the velocity directly by inserting it into
\begin{equation}\label{eq:appendix_polymerization_V}
V=\overline{v_p} + \frac{\epsilon}{2\mathcal{L}}\ \Big( I(u_\mathrm{on},\ u_\mathrm{off}) +  I\big(1-u_\mathrm{on},\ 1-u_\mathrm{off}\big) \Big),
\end{equation}
where $\overline{v_p}=(v_p^++v_p^-)/2$ is the average polymerization velocity.

\bibliographystyle{apsrev4-1}

\end{document}